\documentclass{aa}

\usepackage{graphicx}
\usepackage{txfonts}
\usepackage{lipsum}
\usepackage{lscape}
\usepackage{placeins}

\usepackage{color}
\usepackage{multirow}
\usepackage{array}
\usepackage{natbib}
\bibpunct{(}{)}{;}{a}{}{,}
\usepackage{xcolor}
\usepackage{float}
\usepackage{url}
\usepackage{makecell}
                                
\usepackage[colorlinks=true, 
            linkcolor=blue,
            citecolor=blue,
            urlcolor=blue]{hyperref}

\renewcommand{\makeLineNumber}{\relax}

\begin{document} 

\title{Imaging the positron annihilation line\\with 20 years of INTEGRAL/SPI observations}

   \subtitle{}

   \author{Hiroki Yoneda\inst{1,2,3,4,5},
           Thomas Siegert\inst{1},  
           Saurabh Mittal\inst{1}
          }

   \institute{Julius-Maximilians-Universit\"{a}t W\"{u}rzburg, Fakult\"{a}t f\"{u}r Physik und Astronomie, Institut f\"{u}r Theoretische Physik und Astrophysik, Lehrstuhl f\"{u}r Astronomie, Emil-Fischer-Str. 31, D-97074 W\"{u}rzburg, Germany\\
            \email{hiroki.yoneda.phys@gmail.com}
            \and The Hakubi Center for Advanced Research, Kyoto University, Yoshida Ushinomiyacho, Sakyo-ku, Kyoto 606-8501, Japan
            \and Department of Physics, Kyoto University, Kitashirakawa Oiwake-cho, Sakyo-ku, Kyoto 606-8502, Japan
            \and RIKEN Nishina Center, 2-1 Hirosawa, Wako, Saitama 351-0198, Japan
            \and Kavli Institute for the Physics and Mathematics of the Universe (WPI), UTIAS, The University of Tokyo, 5-1-5 Kashiwanoha, Kashiwa, Chiba 277-8583, Japan
            }
            
   \date{}
   
   \titlerunning{Imaging the positron annihilation line\\with 20 years of INTEGRAL/SPI observations}
   \authorrunning{H. Yoneda et al.}

\abstract
{The electron-positron annihilation line at 511 keV provides a unique probe for studying the distribution and origin of positrons in our Galaxy. The SPI spectrometer on board the International Gamma-Ray Astrophysics Laboratory
(INTEGRAL) has observed this gamma-ray line for two decades.}
{We analyzed 20 years of INTEGRAL/SPI observations to produce the most sensitive all-sky map of the 511 keV line emission to date with the aim of revealing new features and providing refined measurements of known sources.}
{We performed image deconvolution using the Richardson-Lucy algorithm and employed a bootstrap analysis to evaluate statistical uncertainties of fluxes from regions of interest. Systematic uncertainties in the parameter choices for the image reconstruction were also considered. We utilized GPU acceleration to enable this computationally intensive data analysis.}
{The reconstructed image successfully recovers the basic morphological features reported in model-fitting studies: a bright central component, a broad bulge, and an elongated disk component along the Galactic plane.
We also report hints of new spatial features in the reconstructed image, including an asymmetric structure in the broad bulge emission and a 511 keV emission potentially associated with massive stars from the Scorpius–Centaurus and other OB associations.
While the significance of these new features is marginal ($\sim 2\sigma$), they are spatially consistent with $^{26}$Al emission from massive stars in that region, suggesting that this 511 keV emission originates from its $\beta^{+}$ decay.}
{Our 20-year dataset provides the most detailed 511 keV emission map to date, reproducing the global structures suggested in the model-fitting approach while also revealing hints of new spatial features. 
These findings provide new insights into the origin of Galactic positrons and the propagation of low-energy positrons in the interstellar medium.
Future MeV gamma-ray observations, such as with COSI, are expected to confirm the reported features and shed further light on the nature of positrons in our Galaxy.}


   \maketitle

\section{Introduction}

Positrons, the antimatter counterparts of electrons, play fundamental roles in high-energy astrophysics. Their existence was first predicted theoretically by \cite{Dirac1931} and experimentally discovered by \cite{Anderson1932}. When a positron encounters an electron, they can either directly annihilate to produce two photons with 511\,keV energy each if the particles are sufficiently cooled or form a bound state called positronium (Ps). Ps eventually decays, producing two 511\,keV photons (para-Ps) or three photons with a continuum spectrum below 511 keV (ortho-Ps), depending on its spin states. This characteristic spectral signature provides a unique probe of antimatter in the Universe. Understanding the production, propagation, and annihilation of positrons is crucial for studying various astrophysical phenomena, from stellar evolution to particle astrophysics.

The discovery of the Galactic 511 keV emission dates back to balloon experiments \citep[e.g.,][]{Haymes1969,Johnson1973}.
Since then, it has been measured by multiple satellite missions for over 50 years.
OSSE aboard the Compton Gamma Ray Observatory \citep{OSSE} provided the first maps of the 511 keV emission, revealing its diffuse nature at the Galactic Center \citep[e.g.,][]{Purcell1997}.
After that, INTEGRAL/SPI \citep{Winkler2003_INTEGRAL,Vedrenne2003_SPI} revolutionized our understanding with its high energy and angular resolution by unveiling its complex morphology with increasing exposure time since its launch in 2002.

Several key observational features have been obtained from these studies \citep[e.g.,][]{CHURAZOV2020,Prantzos2011}.
Around the Galactic center, the 511 keV emission is dominated by a diffuse bulge component with an extent of 3-12 degrees and is potentially accompanied by a central point source \citep{Skinner2015}.
INTEGRAL/SPI has also detected extended disk emission along the Galactic plane.
The latest measurements \citep{Skinner2015,Siegert2016} fitted data with analytical spatial models for these components and reported the fluxes of the bulge, central point source, and the disk to be $(8.9-10.1)\times10^{-4}$, $(0.6-1.2)\times10^{-4}$, and $(13.1-20.1)\times10^{-4}~\mathrm{ph~cm^{-2}~s^{-1}}$, respectively. 
The ratio of the 511 keV emission from the bulge is comparable to that from the disk emission.
Such a large bulge-to-disk ratio is a unique feature not seen in other wavelengths.
We note that \cite{Skinner2015} claimed that the Galactic disk emission has a scale height of 3$^{\circ}$, which is the $\sigma$ parameter of the two-dimensional Gaussian function for the disk component vertical to the Galactic plane.
In contrast, \cite{Siegert2016} reported that it has a larger scale height of 10.5$^{\circ}$.
We refer to the former and latter as the thin and thick disk models, respectively.
Such a difference is probably caused by different data and background treatments.

Despite decades of study, the origin of the Galactic positrons remains one of the long-term mysteries in high-energy astrophysics, the so-called positron puzzle \citep{Siegert2023}.
Various sources have been proposed to explain the Galactic 511 keV emission, each with a different spatial distribution, for example, $\beta^{+}$ decay of radioisotopes emitted from massive stars \citep{Diehl2006,Diehl2010,Pleintinger2023}, core-collapse supernovae \citep{Siegert2015}, type Ia supernovae \citep{Milne1999,Prantzos2006,Siegert2017PhD}, and classical novae \citep{Leung2022}; e$^{-}$e$^{+}$ production in X-ray binaries and pulsars \citep{Guessoum2006,Weidenspointner2008,Wang2006};
e$^{-}$e$^{+}$ production near the hot inner disk of Sgr A$^{*}$ (for the bulge emission, \citealt{Totani2006}); light dark matter models as an exotic scenario \citep{Finkbeiner2007,Siegert2022,Sheng2024,Watanabe2025}.
Through cosmic-ray interactions, positrons can also be produced by nearby sources such as the Moon and asteroids \citep{Moskalenko2008,Siegert2024,Fujiwara2025}.
However, no single scenario fully explains the observed flux and spatial distributions, suggesting the 511 keV line likely originates from a combination of (some of) these sources.

One of the key approaches to solving the positron puzzle is to study the morphology of the 511 keV emission, as different source candidates predict distinct spatial distributions.
However, direct imaging is not possible in the MeV band, requiring sophisticated image reconstruction techniques that deconvolve the detector response from photon count data.
The first detailed 511 keV map was derived by CGRO/OSSE \citep{Purcell1997}, followed by early INTEGRAL/SPI analysis \citep{Knoedlseder2005}.
Subsequently, \citet{Bouchet2010} performed image analysis using 6 years of SPI data and produced the 511 keV map with $5^\circ \times 5^\circ$ pixel resolution.

In this work, we analyze the spatial features using 20 years of INTEGRAL/SPI observations of the 511 keV emission line, representing the highest available photon statistics. 
We employed the Richardson-Lucy algorithm for image reconstruction, which is particularly suited for recovering the 511 keV map from the complex data and response of the coded-masked telescope of SPI.
Additionally, we optimized the background components simultaneously during the image reconstruction process.
Furthermore, we implemented bootstrap methods to quantify the significance of detected spatial structures and their flux uncertainties.
Such an analysis has not been performed previously due to the heavy computation required. Here, we use GPU-accelerated computing to achieve reasonable uncertainty estimates.

The paper is organized as follows.
In Sect.~\ref{sec_obs_data_reduction}, we describe the SPI observations and data reduction. 
Section~\ref{sec_analysis_methods} presents our image deconvolution methods, background modeling, and uncertainty evaluations. 
Section~\ref{sec_results} presents the updated 511 keV map, revealing potential new spatial features. 
Section~\ref{sec_discussion} discusses these findings with possible interpretations and future prospects with upcoming MeV gamma-ray telescopes.
Finally, we conclude our work in Sect.~\ref{sec_conclusion}.

\section{Observations and data reduction}
\label{sec_obs_data_reduction}

The INTErnational Gamma-Ray Astrophysics Laboratory \citep[INTEGRAL][]{Winkler2003_INTEGRAL} has been observing the gamma-ray sky since October 17, 2002, and ceased its scientific operations on February 28, 2025.
The spectrometer SPI \citep{Vedrenne2003_SPI} on board the satellite was the most sensitive gamma-ray spectrometer telescope with a coded aperture mask ever built.
It measured photons in the energy range from 18\,keV to beyond 8\,MeV with an unprecedented energy resolution of 0.41\% (full width at half maximum; FWHM) at 511\,keV and an angular resolution of $2.7^\circ$ (FWHM) \citep{Diehl2018}.
Due to cosmic-ray bombardment, about 99\% of the recorded events are typically due to instrumental background.
Additionally, this radiation dosage from cosmic rays and trapped particles in the Van Allen radiation belts degrades the energy resolution of the 19 SPI's germanium detectors.
With annealing procedures, heating up the detectors to above $100^\circ\mathrm{C}$ about every half year restored the resolution almost completely \citep{Roques2003}, except for a long-term drift that worsened the resolution over the 22 mission years by about 10\% to 0.46\% at 511\,keV.

In this work, we analyze 20 years of INTEGRAL/SPI data covering the orbital revolutions 43--2679.
It spans from February 19, 2003, to August 29, 2023 (ISDC Julian Date 1145--8641).
Using the ISDC standard tools within its OSA package,\footnote{\url{https://www.astro.unige.ch/integral/analysis\#Software}} we selected pointings according to several selection criteria:
We required that each pointing is at least 120\,s long.
Further, we excluded five hours before and after solar flares, and individual high count rates due to flaring sources or particle events in the detectors.
More details are described in Appendix~\ref{sec_data_selection}.
This left us with 143,833 pointings.
Using a first-order maximum likelihood fit using a background model and a sky model, we found that 206 individual pointings show an erratic behavior in that the background plus sky combination does not describe the data properly, showing residuals with $\pm 10\sigma$ or above.
We filtered out these pointings and performed an additional maximum likelihood with the same background and sky models, which resulted in a screened dataset.
After the data screening, we obtained 143,627 pointing observations with a total gross exposure time of 367.4 Ms.
The typical exposure per observation ranges from $\sim$2 ks to $\sim$4 ks, with some long staring observations exceeding 20 ks. The screened dataset yields an average exposure of 2.56 ks per observation.

\begin{figure}
    \centering
    \includegraphics[width=0.8\linewidth]{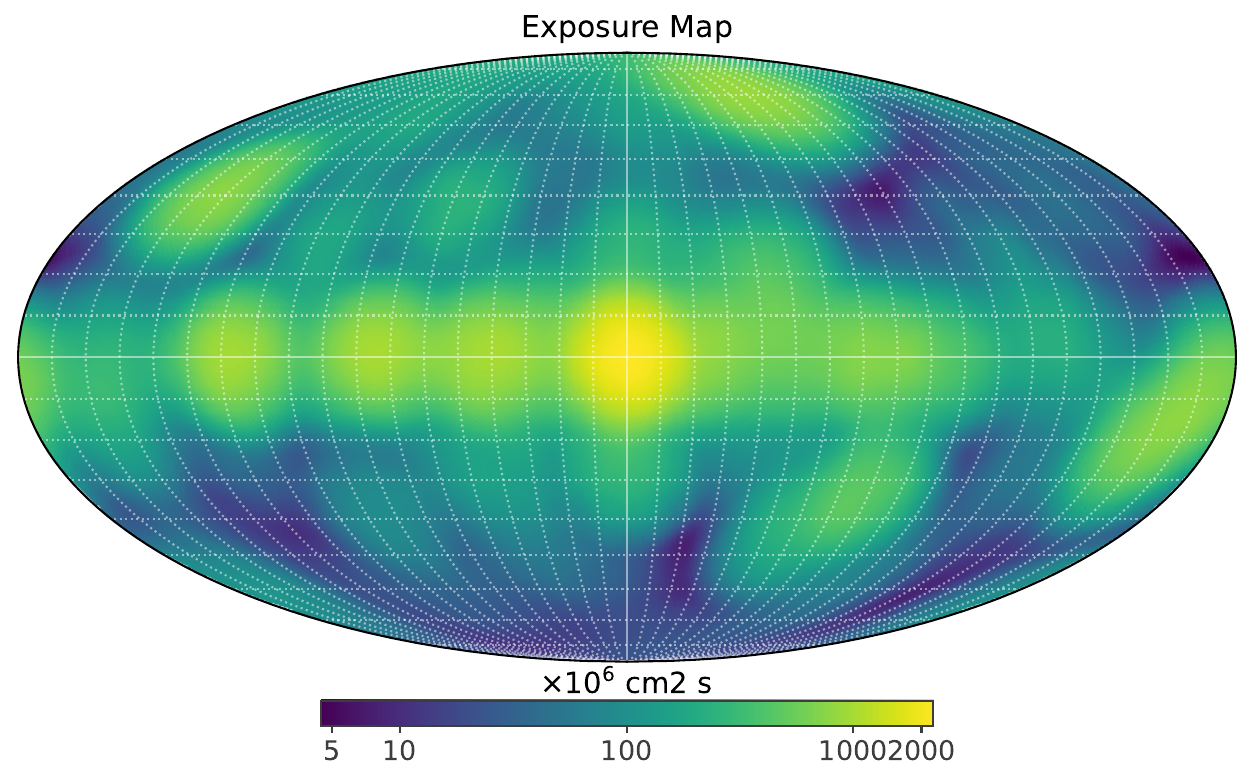}
    \caption{Exposure map for the 511 keV line observations with INTEGRAL/SPI over 20 years in the Galactic coordinates. }
    \label{fig_exp_map}
\end{figure}

\begin{figure}
    \centering
    \includegraphics[width=0.8\linewidth]{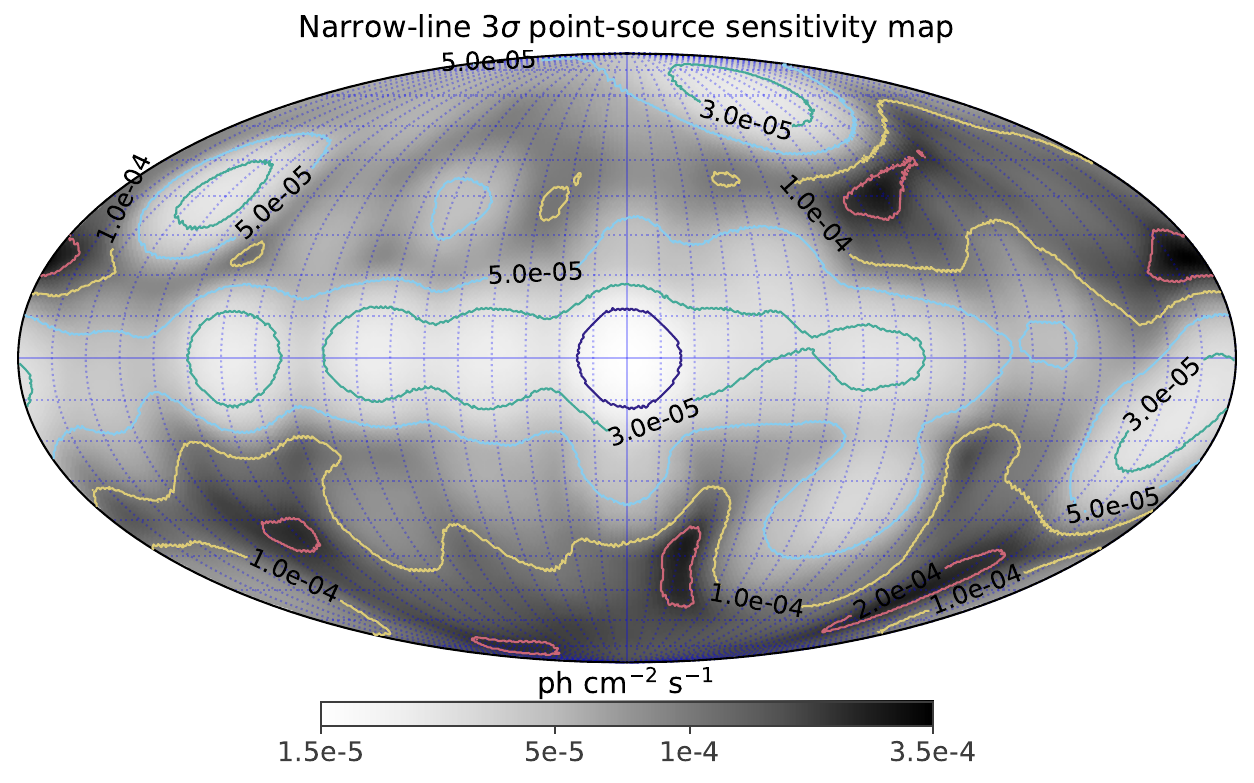}
    \caption{Map of the narrow-line 3$\sigma$ point-source sensitivity at 511 keV derived from the exposure map. The contour lines show the sensitivity of $2\times10^{-5}$, $3\times10^{-5}$, $5\times10^{-5}$, $1\times10^{-4}$, and $2\times10^{-4}~\mathrm{cm^{-2}~s^{-1}}$ from purple to red.}
    \label{fig_sensitivity_map}
\end{figure}

Figure~\ref{fig_exp_map} shows the effective exposure map over the sky derived by the exposure time with the SPI instrument response.
The maximum exposure appears around the Galactic center with $2.2 \times 10^{9}~\mathrm{cm^{2}~s}$, which is about ten times larger than in the first image deconvolution analysis of 511 keV SPI data \citep{Knoedlseder2005}.
Given the narrow-line 3$\sigma$ point-source sensitivity of $5.0 \times 10^{-5}~\mathrm{cm^{-2}~s^{-1}}$ for an exposure of $2.1 \times 10^{8}~\mathrm{cm^{2}~s}$ \citep{Knoedlseder2005},
we converted the exposure map into a point source sensitivity map. 
We note that four detector failures occurred throughout the operation,\footnote{\url{https://integral.esac.esa.int/AODocumentation/SPI_ObsMan.pdf}} reducing the effective area by approximately 20\%.
This reduction is not accounted for in this simple sensitivity estimation, as it does not affect the conclusions of our discussion.
As shown in Fig.~\ref{fig_sensitivity_map}, the point source sensitivity reaches $\sim 3\times 10^{-5}~\mathrm{cm^{-2}~s^{-1}}$ along the Galactic plane.
We also note that the point source sensitivity shown here cannot be applied directly to extended emission, which we discuss in this work.
Our approach to estimate the sensitivity and significance of the extended emission is described in Sect.~\ref{sec_bootstrap}.

With such a long exposure, continuum emissions from bright point sources can contaminate our analysis even within a narrow energy band for this gamma-ray line analysis.
We account for this by modeling and subtracting these sources as ``background'' components.
The exposure is not uniformly distributed in high latitude regions, for example, $|b| > 40^{\circ}$, which may cause artifacts in the image reconstruction discussed later.
We also want to mention that upcoming MeV satellites such as COSI \citep{COSIofficial} will address this issue through all-sky surveys with a wide field-of-view (FoV) Compton telescope.

Our analysis uses only single-detector events (SEs) because they dominate the 511 keV signal data, and the contribution from multiple-detector events (MEs) can be negligible.
We note that SEs are the events that deposit their energy in a single detector, while MEs are those that deposit their energy in more than one detector. More details are found in \cite{Vedrenne2003_SPI}.
The data were analyzed by sorting the events in a three-dimensional data space, spanned by the (calibrated) event energy, the detector number, and the SPI pointing number (see Sect.\,\ref{sec:data_structure}).
An initial energy binning of 0.5\,keV was used by the standard pipeline from ISDC, well below the instrumental energy resolution.
Our final imaging analysis was carried out in a single 6\,keV wide energy bin from 508--514\,keV.
For the analysis of point sources and their contribution to the 511\,keV bin, we used a spectral range from 400--640\,keV with 40 energy bins of 6 keV width each.
In this way, we estimated the continuum flux of different sources (see Sect.\,\ref{sec:background_modeling}).

\section{Analysis methods}
\label{sec_analysis_methods}

\subsection{Data structure}\label{sec:data_structure}

Here, we outline the data structure in our image analysis.
This analysis uses detected counts per energy bin per pointing in each SPI's germanium detector.
We represent it as $D_{p,d,e}$, where $p$, $d$, and $e$ are the indices of the pointings, the detector ID, and the energy bin in the data space, respectively.
We aim to reconstruct the flux map of the 511 keV line in the Galactic coordinates, which we represent as $\lambda_{j}$.
The index $j$ represents the pixel index of the map, and the HEALPix scheme is adopted to describe the 511 keV map \citep{Healpix}.
We choose the pixel size as $\sim 1^{\circ}$, corresponding to the NSIDE parameter of 64, considering SPI's angular resolution of 2.7$^{\circ}$ FWHM.

\subsubsection{Response}
In the image deconvolution, we calculate expected counts in the detectors from the 511 keV map and compare them with the observed data.
For the expected count calculation from the assumed 511 keV map, we first convert it into the flux map in SPI's local coordinates.
Considering the FoV of SPI, which is $16^{\circ} \times 16^{\circ}$ for a fully coded region and $34^{\circ} \times 34^{\circ}$ including partially coded regions, the converted map covers $47^{\circ} \times 47^{\circ}$ centered at the pointing direction of the SPI instrument, sufficiently larger than the instrumental FoV.
We prepare the matrix $C_{p,j,k}$ for each pointing to perform this coordinate transformation.
Here, $k$ denotes the pixel index in the converted map.
The converted map in the local coordinates at the pointing $p$ can be calculated as $\sum_{j} C_{p,j,k} \lambda_{j}$.
Since most of the elements in $C_{p,j,k}$ are zero, we implement $C_{p,j,k}$ in the compressed sparse row (CSR) format as a sparse matrix, which can significantly reduce the memory usage and computational cost by storing only non-zero elements.

Next, we calculate the total expected counts in each detector.
The response of the coded mask aperture is modeled as the matrix $R^{\mathrm{mask}}_{k,d}$, representing the effective area of detector $d$ for a source at pixel $k$.
With the livetime $T_{p,d}$ of each detector and each pointing, the total expected counts can be calculated as $T_{p,d} \sum_{k} \sum_{j} R^{\mathrm{mask}}_{k,d} C_{p,j,k} \lambda_{j}$.
Finally, we consider the response of the germanium detectors. 
The energy redistribution matrix $R^{\mathrm{rmf}}_{e}$ represents the probability that a detected 511 keV gamma ray is measured in energy bin $e$.
Consequently, the expected count $\epsilon^{\mathrm{source}}_{p,d,e}$ at each pointing, detector, and energy bin from a given 511 keV map can be calculated as
\begin{eqnarray}
\label{eq_epsilon_source}
\epsilon^{\mathrm{source}}_{p,d,e} &=& T_{p,d} \sum_{k} \sum_{j} R^{\mathrm{rmf}}_{e} R^{\mathrm{mask}}_{k,d} C_{p,j,k} \lambda_{j}\\
&=& \sum_{j} R_{p,d,e,j} \lambda_{j}~,
\end{eqnarray}
where for simplicity we define $R_{p,d,e,j}$ as
\begin{eqnarray}
R_{p,d,e,j} \equiv T_{p,d} \sum_{k} R^{\mathrm{rmf}}_{e} R^{\mathrm{mask}}_{k,d} C_{p,j,k}~.
\end{eqnarray}

Our analysis utilizes a single energy bin in the data space, spanning from 508 keV to 514 keV.
Since we use SE data, the number of bins for $d$ is 19, which is the same as the number of the germanium detectors of SPI.
The response matrices are generated based on the Python-based INTEGRAL/SPI data analysis package PYSPI \citep{Biltzinger2022}.
The image reconstruction described below is performed using an image analysis framework implemented in the cosipy library \citep{MC2023,Yoneda2025}.

\subsubsection{Background modeling}\label{sec:background_modeling}

The instrumental background in SPI is the most dominant component at all energies except for gamma-ray bursts or bright transients.
We used the method by \citet{Siegert2019}, who defined the instrumental response and background database \citep{Diehl2018}, which was maintained until the end of the INTEGRAL mission.
This database uses the raw count spectra per detector and revolution in 0.5\,keV binning and fits all the spectral features using about 600 gamma-ray lines on top of a multiple broken power-law continuum from 20\,keV to 2\,MeV.
At SPI's high energy range (2--8\,MeV), there are about 1000 more gamma-ray lines to be evaluated, but which cannot be done with high accuracy on the orbital time scale of three days.
Around the 511\,keV line, and in particular for the energy bin of interest from 508--514\,keV, there are two gamma-ray lines on top of a continuum.
Since the raw spectra are dominated by instrumental background, it was found that the relative count rate of different background components in each detector stays constant over time.
These ``detector patterns'' are typically combined to contributions from instrumental lines ($B^{\mathrm{line}}_{p,d,e} $) and instrumental continuum ($B^{\mathrm{cont}}_{p,d,e}$) to avoid over-fitting.

Because the relative variation as a function of time is strictly not predictable, we use a ``tracer function'' in time that estimates the pointing-to-pointing variation to first order well.
For the purpose of the 511 keV line, it was found that the SSATOTRATE (side shield assembly total rate of the SPI anticoincidence shield) traces the 511\,keV background best \citep{Siegert2016,Siegert2019}.
However, there is additional variation that may be related to radionuclide production when INTEGRAL passes the radiation belts.
For this reason, we introduce for both background components (lines $b^{\mathrm{line}}_{\mathrm{rev}(p)}$ and continuum $b^{\mathrm{cont}}_{\mathrm{rev}(p)}$) one re-scaling parameter per revolution.
This finally amounts to $2278 \times 2 = 4556$ background amplitudes, which are optimized simultaneously with the reconstructed image.

For the potentially bright point sources, we fitted the data from 400 to 640\,keV with power-law functions and estimated the integrated flux from 508--514\,keV.
The point sources are included as background components to only show the truly diffuse emission, and are listed with the estimated fluxes in Table~\ref{tab_bright_source_list}.
Although some sources are known as time-variable sources, considering their time-variability in our analysis did not affect the results. Thus, we used the time-averaged flux in this background modeling.

The total expected counts for our background reads
\begin{eqnarray}
\label{eq_epsilon_bkg}
\epsilon^{\mathrm{bkg}}_{p,d,e} = b^{\mathrm{line}}_{\mathrm{rev}(p)} \sum_{p,d,e} B^{\mathrm{line}}_{p,d,e} + b^{\mathrm{cont}}_{\mathrm{rev}(p)} \sum_{p,d,e} B^{\mathrm{cont}}_{p,d,e} + \sum_{m} B^{\mathrm{source},m}_{p,d,e}~.
\end{eqnarray}

Here, the normalization factors $b^{\mathrm{line}}_{\mathrm{rev}(p)}$ and $b^{\mathrm{cont}}_{\mathrm{rev}(p)}$ for the line and continuum backgrounds, respectively, are free parameters that will be optimized in the image analysis.
The last term accounts for contamination from bright sources, whose normalization factors are fixed since they cannot be uniquely determined due to their coupling with the source fluxes at corresponding pixel positions.
The index $m$ represents the label for the considered point sources.

\begin{table}[htbp]
    \centering
    \caption{List of bright sources subtracted in the image reconstruction.} 
    \begin{tabular}{c|cc|c}
        \hline
        Name & \multicolumn{2}{c|}{Location} & Flux\\
        & $l$ (deg.) & $b$ (deg) & ($\times 10^{-5}$ cm$^{-2}$ s$^{-1}$) \\
        \hline
        Crab & 184.55 & -5.78 & 11.08 \\
        Cygnus X-1 & 71.33 & 3.07 & 3.54 \\
        SWIFT J1753.5-0127 & 24.90 & 12.19 & 1.44 \\
        PSR B1509-58 & 320.31 & -1.15 & 0.83 \\
        GRS 1915+105 & 45.37 & -0.22 & 0.71 \\
        Centaurus A & 309.52 & 19.42 & 0.62 \\
        3C 273 & 289.95 & 64.36 & 0.43 \\
        \hline
    \end{tabular}
    \tablefoot{The fluxes shown here are integrated over the energy range of 508 keV to 514 keV. The source locations are given in Galactic coordinates.}
    \label{tab_bright_source_list}
\end{table}

\subsubsection{Expected count calculation}

The total expected counts consist of contributions from both the 511 keV sky map and the background components. 
Combining these components, the expected counts can be calculated as
\begin{eqnarray}
\label{eq_Estep}
\begin{split} 
\epsilon_{p,d,e} & = \epsilon^{\mathrm{source}}_{p,d,e} + \epsilon^{\mathrm{bkg}}_{p,d,e} \\
& = \sum_{j} R_{p,d,e,j} \lambda_{j} \\
& + b^{\mathrm{line}}_{\mathrm{rev}(p)} \sum_{p,d,e} B^{\mathrm{line}}_{p,d,e} + b^{\mathrm{cont}}_{\mathrm{rev}(p)} \sum_{p,d,e} B^{\mathrm{cont}}_{p,d,e} + \sum_{m} B^{\mathrm{source},m}_{p,d,e}~.
\end{split} 
\end{eqnarray}

\subsection{Image deconvolution}
\label{sec_RL}

In this paper, we focus on the image reconstruction from SPI data.
For reconstruction of the 511 keV sky map, we employ the Richardson-Lucy (RL) algorithm, which is widely used in gamma-ray astronomy \citep{Richardson,Lucy1974}.
It finds an image that maximizes the following log-likelihood function by iteratively updating the image:
\begin{align}
\log L = \sum_{p,d,e} \left(D_{p,d,e} \log \epsilon_{p,d,e} - \epsilon_{p,d,e}\right)~.
\end{align}
Our implementation follows the methodology previously applied to 511 keV data analysis by \cite{Knoedlseder1999,Knoedlseder2005,Siegert2020} with several modifications.

The RL algorithm iteratively updates the sky map based on the comparison between the observed and expected counts.
In each iteration, the update of the sky map is calculated as
\begin{align}
\delta \lambda_{j} &= \dfrac{\lambda^{\mathrm{old}}_j}{\displaystyle \sum_{p,d,e} R_{p,d,e,j}} \displaystyle \sum_{p,d,e} \left(\dfrac{D_{p,d,e}}{\epsilon_{p,d,e}} -1 \right) R_{p,d,e,j},\\
\lambda^{\mathrm{new}}_j & = \lambda^{\mathrm{old}}_j + \alpha \left[\omega_{j} \delta \lambda_{j}\right]_{\mathrm{smooth}}~,
\end{align}
where $\epsilon_{p,d,e}$ is the expected count calculated by Eq.~\ref{eq_Estep}; $\alpha$ is a parameter to accelerate the RL algorithm.
Here, the weight factor $\omega_{j}$ is introduced to enhance the reconstruction efficiency by providing stronger feedback to regions with a longer exposure:
\begin{align}
\omega_{j} \equiv \left( \frac{\sum_{p,d,e} R_{p,d,e,j}}{\mathrm{max} \left( \sum_{p,d,e} R_{p,d,e,j} \right)} \right)^{l}~.
\end{align}
The parameter $l$ controls the strength of this weighting.
While the previous studies \citep{Knoedlseder2005,Siegert2020} adopted $l=0.5$, we explored various values to determine the optimal choice (see Sect.~\ref{sec_systematics}).
Also, following \cite{Siegert2020}, we adopt the operator $[\cdot]_{\mathrm{smooth}}$ representing Gaussian smoothing applied to the weighted difference map.
This smoothing suppresses high-frequency noise in the image, enhancing stability in the image reconstruction.
However, a too-large smoothing kernel can lead to losing information about high-frequency structures in the image.
We optimized the FWHM of this Gaussian kernel $\theta_{\mathrm{smooth}}$ with the data, which is described in the next subsection.

The acceleration parameter $\alpha$ is determined as the largest value that does not produce negative-value pixels in the updated image, i.e., $\alpha = \mathrm{max} \left( - \lambda^{\mathrm{old}}_{j}/ \left[\omega_{j} \delta \lambda_{j}\right]_{\mathrm{smooth}} \right)$ for $j$ satisfying $\left[\omega_{j} \delta \lambda_{j}\right]_{\mathrm{smooth}} < 0$.
In some cases, we noticed that allowing an excessively large value for $\alpha$ could cause instability in the image reconstruction.
Thus, we also added a constraint that $\alpha$ must be smaller than 100 (see Appendix~\ref{sec_dep_max_alpha} for details).
While \cite{Knoedlseder2005} optimized $\alpha$ by maximizing the likelihood under the non-zero pixel condition,
our implementation focuses on computational simplicity by adopting the maximum possible $\alpha$ that maintains numerical stability.
The difference in these approaches mainly affects the convergence speed rather than the final solution.

Furthermore, we implemented simultaneous optimization of background normalization factors using the RL algorithm:
\begin{align}
b^{\mathrm{line/cont,new}}_{\mathrm{rev}} =  \dfrac{b^{\mathrm{line/cont,old}}_{\mathrm{rev}}}{\displaystyle \sum_{p \in \sigma(\mathrm{rev}),d,e} B^{\mathrm{line/cont}}_{p,d,e}} \displaystyle \sum_{p\in \sigma(\mathrm{rev}),d,e} \dfrac{D_{p,d,e}}{\epsilon_{p,d,e}} B^{\mathrm{line/cont}}_{p,d,e}~,
\end{align}
where $\sigma(\mathrm{rev})$ denotes the set of pointings in a given orbital revolution.
To assess the impact of this background optimization, we also performed reconstructions without updating these normalization factors, which were already optimized during the background modeling and data screening phase.
This alternative approach is used to evaluate our systematic uncertainty, as explained in the following subsection.

The iteration process is terminated when the improvement in the log-likelihood ($\Delta L = \log L^{\mathrm{new}} - \log L^{\mathrm{old}}$) becomes smaller than a threshold.
Here we set it to $10^{-2}$.
It is known that running iterations until the likelihood converges can cause artifacts to appear in the reconstructed image.
Previous studies often stopped the iterations early to focus only on global structures.
However, using the 20-year dataset, we aim to search for new features in addition to well-known large-scale structures.
Therefore, we continued the iterations until reaching reasonable convergence in the likelihood.
To address the concern about the artifacts, we employ the bootstrap analysis to evaluate statistical fluctuations, allowing us to assess the uncertainties and significance of features of interest.
Its details are presented in Sect.~\ref{sec_bootstrap}.

\subsection{Parameter optimization and systematic error estimation}
\label{sec_systematics}

The RL algorithm involves several key parameters that affect the reconstructed image:
the weighting power $l$, the smoothing kernel size $\theta_{\mathrm{smooth}}$, the intensity of an initial flux map $\lambda_{\mathrm{init}}$.
Additionally, the choice of background optimization approach can impact the results.
We performed a systematic study to determine the optimal parameter set and evaluate associated systematic uncertainties.

To find the optimal parameters, we performed the image reconstructions with various parameter combinations and identified one that maximizes the log-likelihood of the reconstructed image.
Regarding the smoothing kernel size, we observed that $\theta_{\mathrm{smooth}}$ of 2 degrees yielded the maximum log-likelihood, which is close to SPI's angular resolution of 2.7 degrees.
The weighting power $l$ shows optimal values between 0.5 and 1.0, with the maximum likelihood achieved at $l = 0.75$.
We note that the previous studies typically adopt $l = 0.5$.
Also, the reconstructed image is affected by $\lambda_{\mathrm{init}}$, which is the intensity of an initial flux map with units of cm$^{-2}$ s$^{-1}$ sr$^{-1}$, as a starting point for the image reconstruction.
A uniform intensity across the sky was assumed as the initial condition.
The choice of $\lambda_{\mathrm{init}}$ influences the results, yielding different total fluxes of the reconstructed images.
Based on comparison with the previous model-fitting studies \citep{Skinner2015,Siegert2016},
we adopted $\lambda_{\mathrm{init}} = 5 \times 10^{-5}$ cm$^{-2}$ s$^{-1}$ sr$^{-1}$ as our optimal value because with this value the reconstructed total flux becomes consistent with the previous measurements best.
The details of the parameter optimization are described in Appendix~\ref{sec_details_parameter_optimization}.

The optimal parameter set determined from the above analysis is summarized in Table~\ref{tab_parameter_list}.
Small deviations from the optimal parameters could potentially affect the reconstructed flux distribution, particularly in regions with lower exposure or faint signals. 
We performed additional image reconstructions to evaluate these systematic uncertainties by varying each parameter around its optimal value, as listed in Table~\ref{tab_parameter_list}. 
We also considered reconstructions with fixed background normalization, which has typically been employed in previous studies, as part of our systematic error evaluation. 
For each region of interest, we calculated fluxes from the reconstructed images with these different parameter sets and used their root mean square to estimate their systematic uncertainty.

We note that additional systematic uncertainties could arise from the separation between the line and continuum components in the background templates.
We evaluated this by perturbing the relative ratio between these two components for each detector and each revolution at the ~2\% level (1$\sigma$), based on the line-to-continuum background ratio variations observed in SPI background behavior \citep{Diehl2018}. 
The resulting flux variations remain below 30\%, primarily in low-flux regions. 
These uncertainties are absorbed within the systematic errors from background optimization approaches (fixed vs. optimized normalization factors).

We also evaluated systematic uncertainties from instrument response limitations, especially imperfections of the coded mask response function. We performed the image reconstruction with a modified response matrix of the coded mask aperture, where elements with effective areas of $<$0.2 cm$^{2}$ (corresponding to the edges and off-axis regions of the coded mask pattern) were set to zero. 
The resulting flux variations are $\lesssim$10\% at most, especially in low-flux regions, and these uncertainties are also absorbed within other systematic errors.
We note that photons penetrating the surrounding BGO shield could contribute a somewhat uniform background across all detectors, especially photons from outside $47^\circ\times47^\circ$, where the mask response is not covered in this analysis. This may potentially lead to flux misattribution. Quantitative evaluation of this effect would require detailed simulations of the entire payload geometry. While this is beyond the scope of this analysis, it should be noted that such systematic uncertainties are not accounted for in our current uncertainty estimation.

\begin{table}[htbp]
    \centering
    \caption{Parameter sets used in the image deconvolution and parameter variations for systematic uncertainty estimation.}
    \begin{tabular}{c|c|c}
        \hline
        Parameter & Optimal value & \makecell{Values tested for\\sys. uncertainty}\\
        \hline
        $l$ & 0.75 & 0.5, 1.0 \\
        $\theta_{\mathrm{smooth}}$ (deg., FWHM) & 2.0 & 1.5, 2.5 \\
        $\lambda_{\mathrm{init}}$ (cm$^{-2}$ s$^{-1}$ sr$^{-1}$) & $5\times10^{-5}$ & $3\times10^{-5}$, $7.5\times10^{-5}$ \\
        \hline
    \end{tabular}
    \label{tab_parameter_list}
\end{table}

\subsection{Bootstrap analysis}
\label{sec_bootstrap}

We performed a bootstrap analysis \citep{Efron1979} using two approaches to evaluate the statistical uncertainties of the fluxes of spatial features in the reconstructed image and their detection significance.

To evaluate flux uncertainties in regions of interest, we generate bootstrap samples by replacing the observed counts $D_{p,d,e}$ with values randomly drawn from Poisson distributions whose expected values are the observed counts themselves: $D_{p,d,e} \leftarrow \mathrm{Possion} \left(D_{p,d,e}\right)$.
This is reasonable because the number of events per bin per detector per pointing is typically around 800, and no bin has zero counts in the dataset when the detectors are active.
Next, we apply the RL algorithm to each resampled dataset and obtain a reconstructed image and a flux from the region of interest.
The statistical uncertainties are then derived as the standard deviation of the flux distribution of these bootstrap samples.

We also generate bootstrap samples under the null hypothesis of no signal to discuss the detection significance of the 511 keV emission in specific regions.
Namely, we replace the observed counts $D_{p,d,e}$ with values randomly drawn from Poisson distributions whose expected values are calculated from the background-only model: $D_{p,d,e} \leftarrow \mathrm{Possion} \left(\epsilon^{\mathrm{bkg}}_{p,d,e}\right)$.
The significance is evaluated by comparing the observed flux with the distribution of fluxes obtained from these background-only bootstrap samples.
The fraction of samples that yield fluxes larger than the observed one is shown as a chance probability in this work.

Here, the background-induced fluctuations are estimated by using the same number of iterations as the reconstruction to be evaluated.
Namely, for both approaches, we terminate the iterations at the same point, the 439th iteration, as used in the original data to ensure consistency in the analysis procedure.
This is particularly important for the background-only bootstrap samples.
If we adopt the convergence criterion described in Sect.~\ref{sec_RL}, they often require more iterations to reconstruct an image with no signals, resulting in an underestimation of the background fluctuations in our case.

\subsection{Computational environment}

We performed our data analysis on a workstation equipped with an Intel Core i9-10980XE processor and 256 GB memory.
To accelerate computational time related to the response matrix operations, especially the coordinate transformation for each pointing, we utilized two NVIDIA A6000 GPUs and the CuPy library \citep{cupy_learningsys2017}. 
With this setup, we were able to execute approximately 30 iterations per minute. 
The image reconstruction with the optimal parameter set took about 20 minutes to complete.

For the bootstrap analysis, we spent several weeks generating $6.6\times10^{2}$ samples for each approach.
The considerable computational time required for the bootstrap analysis constrained us to this moderate number of samples.
However, it is worth noting that this bootstrap analysis became feasible owing to the significant speedup achieved through GPU acceleration in the computation of the image reconstruction.
Such a bootstrap analysis of the image reconstruction has not been performed in previous studies of the 511 keV emission, probably due to its computational intensity.

\section{Results}
\label{sec_results}

\subsection{Global features}

\begin{figure*}[!htbp]
    \centering
    \includegraphics[width=0.95\linewidth]{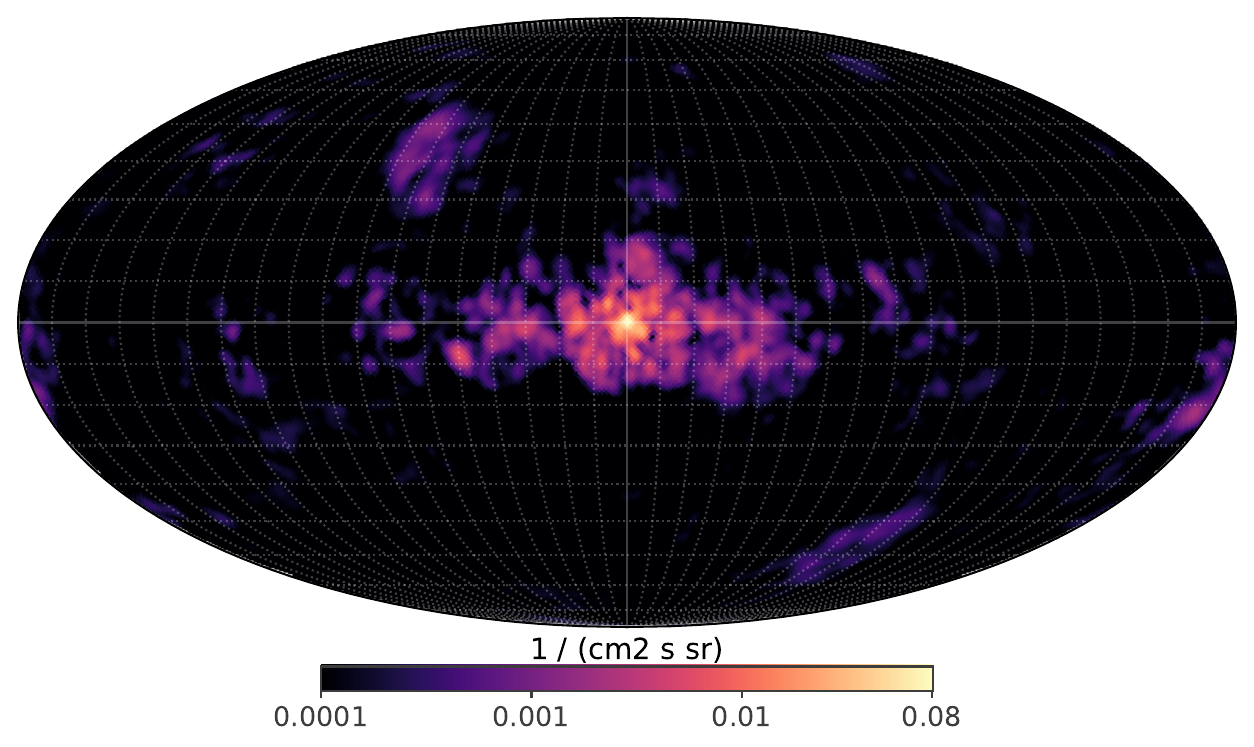}
    \caption{Reconstructed image of the 511 keV line emission from the 20-year INTEGRAL/SPI data. The image is shown in Galactic coordinates. 
    We note that several dark spots appearing around the Galactic center, such as the one at $(l, b) \approx (353^{\circ}, -4^{\circ})$, are artifacts often seen in the image deconvolution.}
    \label{fig_best_image}
\end{figure*}

Figure~\ref{fig_best_image} shows the reconstructed map of the 511 keV line emission from our Galaxy.
The map clearly shows the spatial features suggested by previous model-fitting approaches \citep{Skinner2015,Siegert2016}: a bright central component within $\sim$ 3 degrees of the Galactic center (corresponding to the narrow bulge plus the central point source), an extended bulge emission, and an elongated disk component along the Galactic plane.

The total flux integrated over the bulge region ($|l| < 20^{\circ}$ and $|b| < 20^{\circ}$) is measured to be $(1.36 \pm_{\mathrm{stat}} 0.04 \pm_{\mathrm{sys}} 0.05) \times 10^{-3}$ ph cm$^{-2}$ s$^{-1}$.
The value after $\pm_{\mathrm{stat}}$ represents the statistical error calculated as the standard deviation of the flux distribution extracted from the bootstrapped samples,
and the one after $\pm_{\mathrm{sys}}$ represents the systematic error derived from the results with different parameter sets (see Sect.~\ref{sec_systematics}).
The measured flux shows agreement with the previous model-fitting results, as shown in Table~\ref{tab_global_flux}.
We note that our bulge flux measurement includes the disk emission overlapping with the bulge region, as we directly obtain the flux from the reconstructed image.

When including the Galactic plane, i.e., $|l| < 180^{\circ}$ and $|b| < 20^{\circ}$,
the total flux is derived to be $(2.09 \pm_{\mathrm{stat}} 0.08 \pm_{\mathrm{sys}} 0.23) \times 10^{-3}$ ph cm$^{-2}$ s$^{-1}$.
While this value is slightly lower than that from the model-fitting approaches, it remains consistent within systematic uncertainties. 
We also show the derived flux from the entire region in Table~\ref{tab_global_flux}, which matches well with the fitted models, because $\lambda_{\mathrm{init}}$ was chosen to ensure that the reconstructed total flux falls within the range of previous model-fitting results.

\begin{figure}
    \centering
    \includegraphics[width=0.9\linewidth]{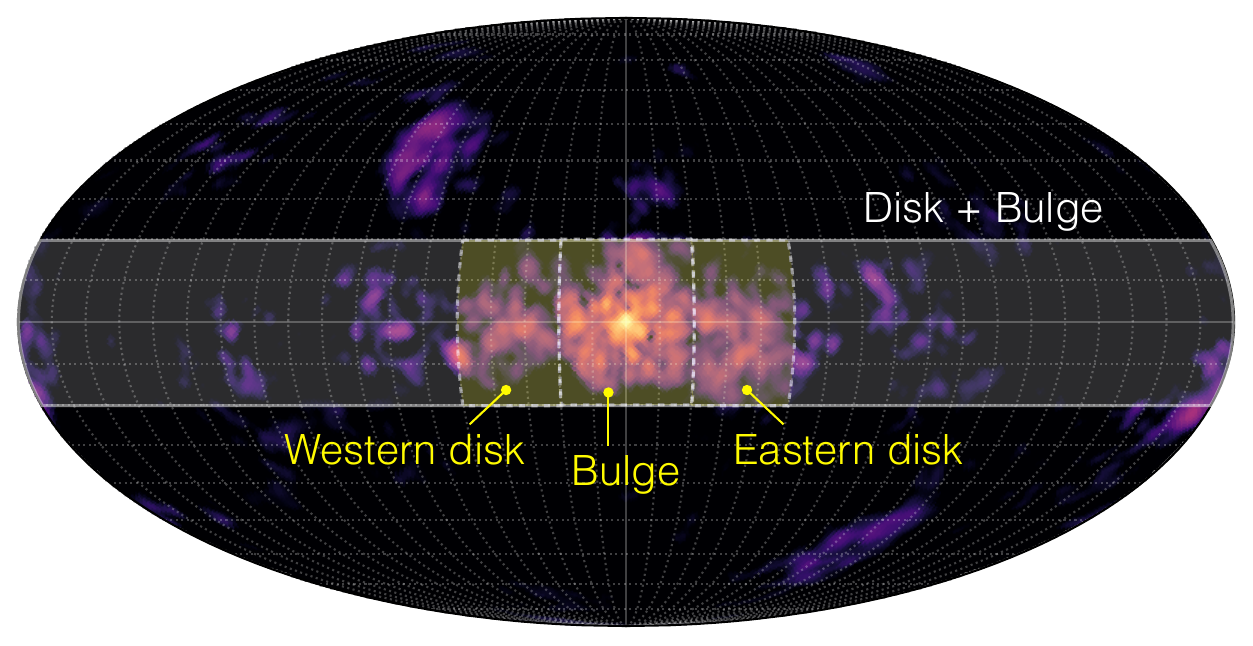}
    \caption{Same as Fig.~\ref{fig_best_image} but with annotations showing the regions used for the flux measurements in Table~\ref{tab_global_flux}.}
    \label{fig_image_annotation_global}
\end{figure}

\begin{table*}[htbp]
    \centering
    \caption{Measured fluxes of the 511 keV line emission for different regions and their comparison with thin and thick disk models.}
    \begin{tabular}{c|cc|c|cc}
        \hline
        Name & \multicolumn{2}{c|}{Region} & Flux & Thick Disk & Thin Disk\\
        & $l$ (deg.) & $b$ (deg.) & \multicolumn{3}{c}{($\times 10^{-3}$ ph cm$^{-2}$ s$^{-1}$)}\\
        \hline
        Bulge & [$-20$, $20$] & [$-20$, $20$] & $1.36 \pm_{\mathrm{stat}} 0.04 \pm_{\mathrm{sys}} 0.05$ & 1.46 & 1.39\\
        Bulge + Disk & [$-180$, $180$] & [$-20$, $20$] & $2.09 \pm_{\mathrm{stat}} 0.08 \pm_{\mathrm{sys}} 0.23$ & 2.62 & 2.51\\
        Eastern Disk & [$310$, $340$] & [$-20$, $20$] & $0.26 \pm_{\mathrm{stat}} 0.04 \pm_{\mathrm{sys}} 0.03$ & 0.27 & 0.18 \\
        Western Disk & [$20$, $50$] & [$-20$, $20$] & $0.19 \pm_{\mathrm{stat}} 0.03 \pm_{\mathrm{sys}} 0.03$ & 0.27 & 0.18 \\
        \hline
        \hline
        All & & & $2.63 \pm_{\mathrm{stat}} 0.08 \pm_{\mathrm{sys}} 0.33$ & 2.72 & 2.52 \\
        \hline
    \end{tabular}
    \label{tab_global_flux}
\end{table*}

\subsection{Longitude and latitude profiles}
\label{sec_profile}

\begin{figure}
    \centering
    \includegraphics[width=0.9\linewidth]{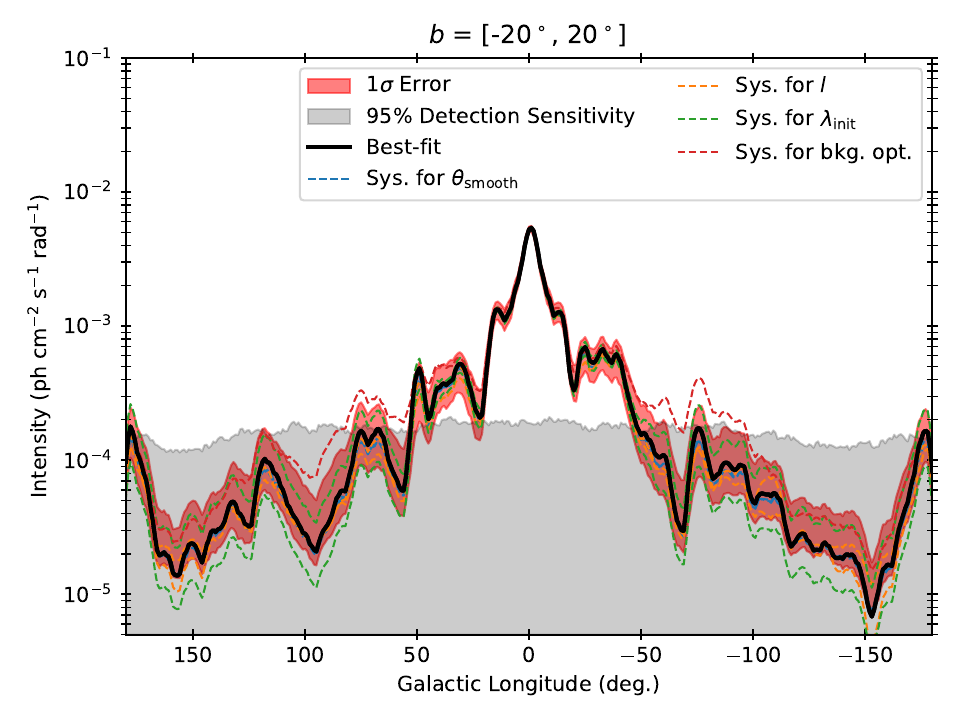}
    \caption{Longitude profile of the 511 keV line emission obtained by integrating the flux from $b = -20$ to $+20$ degrees in 4-degree longitude bins. The red region shows 1$\sigma$ statistical uncertainties derived from bootstrap samples. The dashed lines show systematic uncertainties estimated by varying the parameters in the image reconstruction algorithm. The blue band represents the 95\% detection sensitivity derived from background-only bootstrap samples.}
    \label{fig_profile_longitude}
\end{figure}

To investigate the detailed spatial structure of the 511 keV emission, we examine flux profiles along both Galactic longitude and latitude.
Figure~\ref{fig_profile_longitude} shows the longitude profile obtained by integrating the flux from $b$ = $-20^{\circ}$ to $+20^{\circ}$ in 4-degree longitude bins.
While there is a slight flux enhancement in the negative longitude,
this asymmetry is not statistically significant when considering both statistical and systematic uncertainties.
To quantify this, we measured the fluxes in two specific regions, the eastern and western disk regions, as shown in Fig.~\ref{fig_image_annotation_global}.
The eastern disk ($l$ = $310^{\circ} - 340^{\circ}$) shows the integrated flux of $(0.26 \pm_{\mathrm{stat}} 0.04 \pm_{\mathrm{sys}} 0.03 ) \times 10^{-3}$ ph cm$^{-2}$ s$^{-1}$, 
while the western disk ($l$ = $20^{\circ} - 50^{\circ}$) yields $( 0.19 \pm_{\mathrm{stat}} 0.03 \pm_{\mathrm{sys}} 0.03 ) \times 10^{-3}$ ph cm$^{-2}$ s$^{-1}$.
The former is about 40\% larger than the latter, but the statistical error is also $\sim$30\%.
Thus, the ratio between these regions is consistent with unity within current uncertainties.
These results are also consistent with \cite{Skinner2015,Siegert2016}, rather than the asymmetric profile claimed in early studies \citep{Weidenspointner2008}.

\begin{figure}
    \centering
    \includegraphics[width=0.9\linewidth]{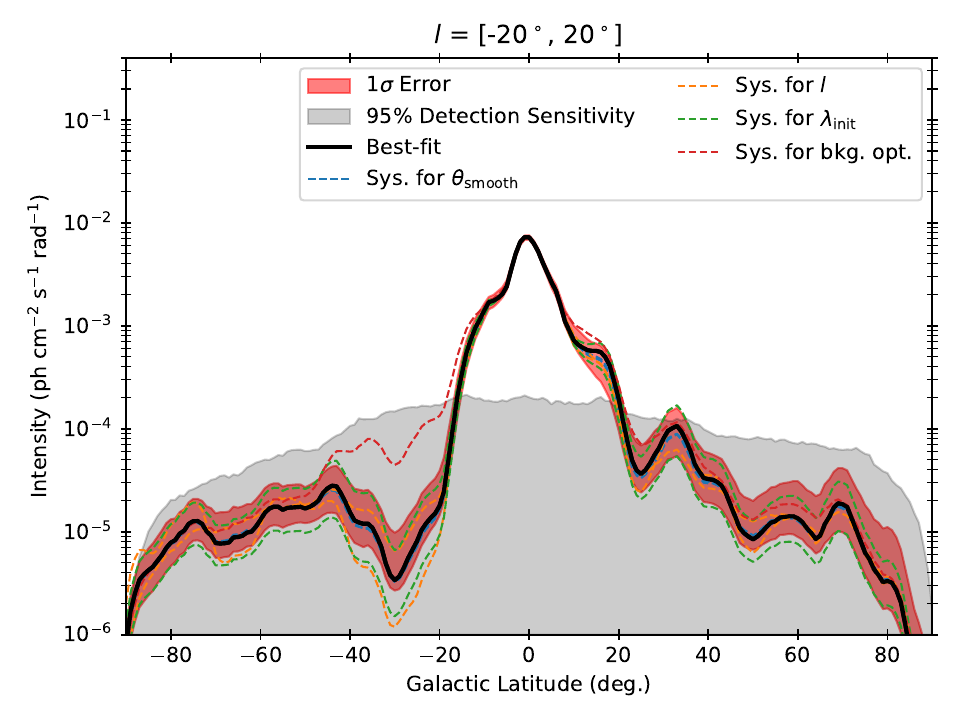}
    \caption{Latitude profile of the 511 keV line emission obtained by integrating the flux from $l = -20$ to $+20$ degrees in 4-degree latitude bins. The uncertainties are shown in the same manner as in Fig.~\ref{fig_profile_longitude}.}
    \label{fig_profile_latitude}
\end{figure}

The latitude profile (Fig.~\ref{fig_profile_latitude}), derived by integrating the flux between $l = -20^{\circ}$ and $+20^{\circ}$ in 4-degree latitude bins, shows potentially interesting features in the vertical structure of the 511 keV emission. The bulge emission exhibits different extensions at $|b| > 10^{\circ}$. While the flux drops sharply at $b \approx -10^{\circ}$ in the northern region, the flux exhibits a plateau between $b = 10^{\circ}$ and $\approx 18^{\circ}$. This corresponds to the more extended structure at ($l$, $b$) = ($-10^{\circ}$--$0^{\circ}$, $10^{\circ}$--$20^{\circ}$) in the northeastern region of the bulge (see also Fig.~\ref{fig_best_image}).

\begin{figure}
    \centering
    \includegraphics[width=0.9\linewidth]{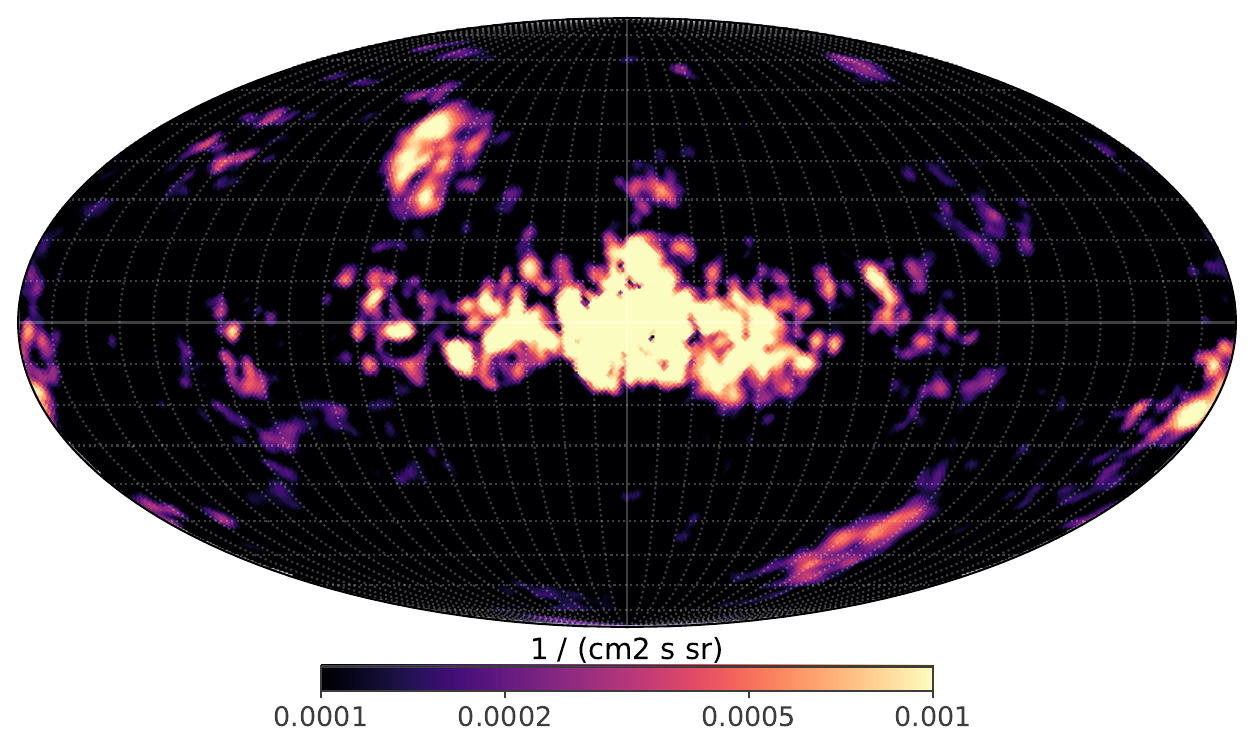}
    \caption{Same as Fig.~\ref{fig_best_image} but with a reduced maximum intensity of $1 \times 10^{-3}$ ph cm$^{-2}$ s$^{-1}$ sr$^{-1}$ to enhance the visibility of faint structures.}
    \label{fig_bestimage_w_max1e-3}
\end{figure}

Figure~\ref{fig_bestimage_w_max1e-3} shows the reconstructed image with a reduced maximum value.
It shows the asymmetric morphology of the Galactic bulge component along the north-south direction more clearly.
The extended bulge emission to the north-east direction is seen as a chimney-like structure, and above it, a blob-like structure can be seen around $(l, b) \approx (-10^{\circ}, 30^{\circ})$.
To discuss this further, we also present the longitude profile with different integration ranges in Appendix~\ref{sec_longitude_profile_more}.

\subsection{Individual regions}

\subsubsection{Search for the 511 keV from the star-forming regions}
\label{sec_search_for_511keV_from_SFR}

\begin{figure}
    \centering
    \includegraphics[width=0.9\linewidth]{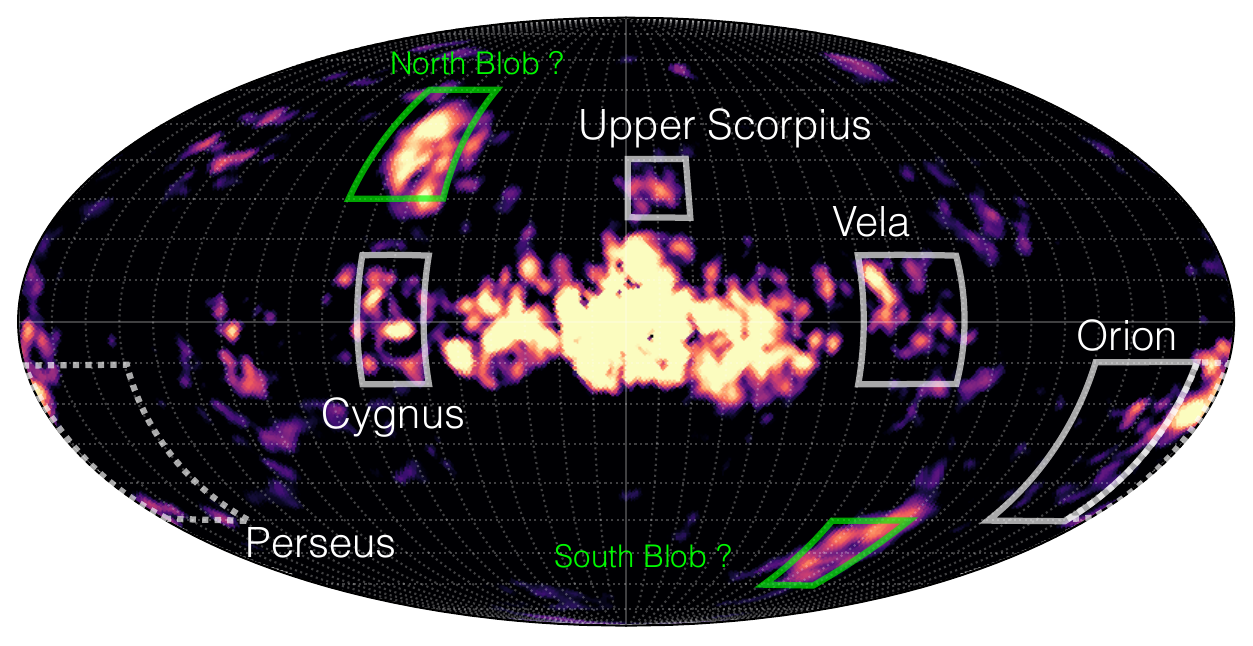}
    \includegraphics[width=0.9\linewidth]{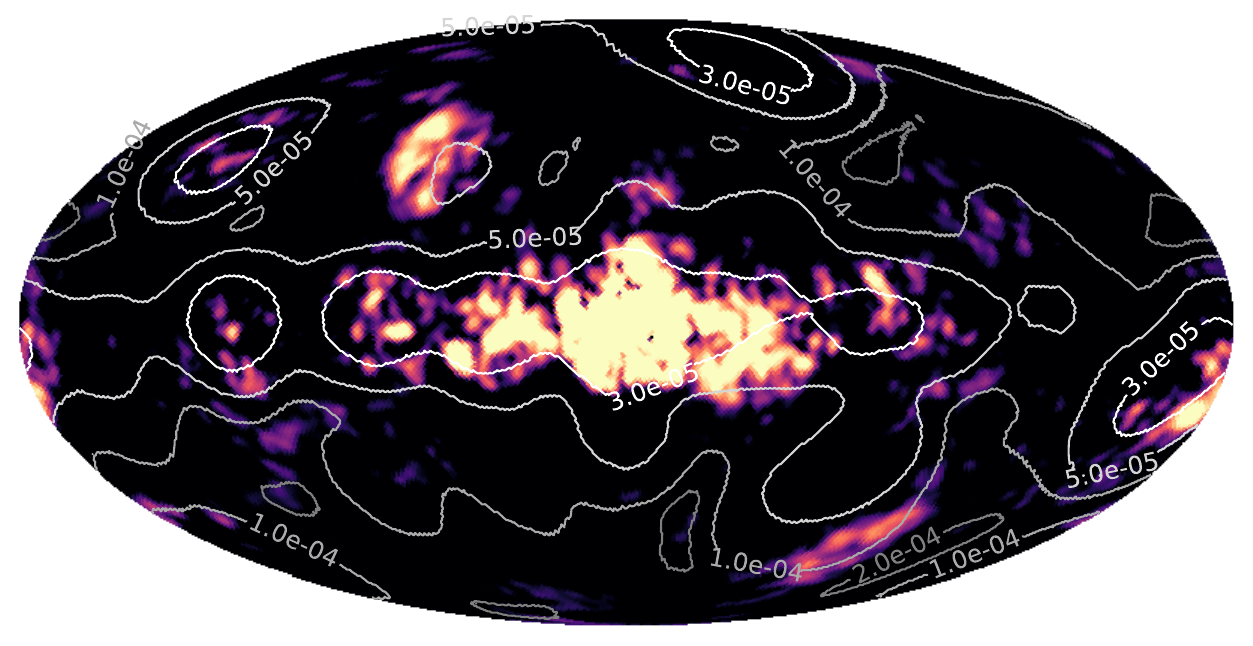}
    \caption{(Top) Same as Fig.~\ref{fig_bestimage_w_max1e-3} but with annotations showing the regions used for flux measurements from individual sources (see Table~\ref{tab_flux_indivisuals}). (Bottom) Same as the top panel but overlaid with contours of the sensitivity map shown in Fig.~\ref{fig_sensitivity_map}.}
    \label{fig_bestimage_w_max1e-3_annotation}
\end{figure}

Here, we investigate the 511 keV emission from 
individual regions, particularly the Cygnus region and the Scorpius–Centaurus OB association (see Fig.~\ref{fig_bestimage_w_max1e-3_annotation}).
Including other star-forming regions, we evaluate the fluxes and their chance probabilities using the bootstrap samples with only the background.

The Cygnus region, one of the nearest complexes of massive star formation, shows a flux of $(4.2 \pm_{\mathrm{stat}} 2.0 \pm_{\mathrm{sys}} 2.0 ) \times 10^{-5}$ ph cm$^{-2}$ s$^{-1}$
at a chance probability of $6.4\times10^{-2}$.
Here, the region is defined as $l = 60^{\circ}$ to $80^{\circ}$ and $b = -15^{\circ}$ to $15^{\circ}$.
Since the significance is about a 2$\sigma$ level only considering statistical uncertainties, it should be interpreted as a hint of the 511 keV emission from the Cygnus region.
Thus, we also show the 99\% upper limit of the flux value in Table~\ref{tab_flux_indivisuals}.
Appendix~\ref{sec_bootstrap_figures} shows the flux distribution derived from the bootstrap samples.
We note that this emission could be attributed to either enhanced positron production from stellar activity in the Cygnus region or detection of the Galactic disk component due to deep exposure in this region. 
The expected flux from the disk component is estimated as $\lesssim 10^{-4}$ ph cm$^{-2}$ s$^{-1}$, making it difficult to distinguish between these two contributions with current analysis conclusively, which is discussed more in Sect.~\ref{sec_discussion} along with future MeV gamma-ray observations.

The Upper Scorpius region ($l = 340^{\circ}$ to $360^{\circ}$, $b = 25^{\circ}$ to $40^{\circ}$) is a part of the Scorpius–Centaurus association, the nearest OB association to the Sun.
It shows a flux of $(1.4 \pm_{\mathrm{stat}} 0.8 \pm_{\mathrm{sys}} 0.5) \times 10^{-5}$ ph cm$^{-2}$ s$^{-1}$ and a chance probability of $8.0 \times 10^{-2}$.
This region corresponds to the chimney-like structure discussed in the previous subsection.
Interestingly, $^{26}$Al maps measured by COMPTEL and INTEGRAL/SPI also show an enhanced emission from this region \citep{Pluschke2001,Diehl2010}.
These findings suggest a possible spatial correlation between $^{26}$Al and positron annihilation.
Considering that $^{26}$Al also emits a positron via its $\beta^{+}$ decay process,
this correlation might shed light on the asymmetric vertical structure and the origin of the 511 keV emissions in this region with further confirmation.
It is discussed more in Sect.~\ref{sec_discussion}, also regarding the ``OSSE 511\,keV fountain.''

We also investigated the 511 keV emission from other prominent star-forming regions, the Vela, Orion-Eridanus, Perseus, and Orion regions.
The Vela region ($l = 260^{\circ}$ to $290^{\circ}$, $b = -15^{\circ}$ to $15^{\circ}$) shows a large chance probability of 19\% with an upper limit of $< 8.3 \times 10^{-5}$ ph cm$^{-2}$ s$^{-1}$.
The Orion region ($l = 190^{\circ}$ to $220^{\circ}$, $b = -50^{\circ}$ to $-10^{\circ}$) similarly shows no significant excess when considered alone.
When extending the region to include the Perseus OB association, the chance probability is somewhat improved to 10.1\%.
It should be noted that there could be possible contamination from the statistical fluctuation of the nearby bright source, the Crab.
When we extracted the bright region ($l = 170^{\circ}$ to $200^{\circ}$, $b = -30^{\circ}$ to $0^{\circ}$), its flux was derived to be $6.5 \pm_{\mathrm{stat}} 2.4 \pm_{\mathrm{sys}} 1.9$ ph cm$^{-2}$ s$^{-1}$ with an improved chance probability of $1.2 \times 10^{-2}$.
A faint signal may be seen at $l = 120^{\circ}$ close to the Galactic plane in Fig.~\ref{fig_bestimage_w_max1e-3_annotation}, but its chance probability is about 60\%, and we interpret it as a fluctuation in the reconstructed image.

\subsubsection{Other regions}

The reconstructed image shows a blob-like structure at high Galactic latitude around ($l$, $b$) = (70$^{\circ}$, 40$^{\circ}$).
When defining the region as $l = 60^{\circ}$ to $90^{\circ}$, $b = 30^{\circ}$ to $60^{\circ}$,
we measured the flux of this region to be $(7.5 \pm_{\mathrm{stat}} 2.4 \pm_{\mathrm{sys}} 3.4) \times 10^{-5}$ ph cm$^{-2}$ s$^{-1}$.
None of the background-only bootstrap samples reproduced such a high flux in this region.
Since no known bright high-energy source has been identified in this region,
it could be an interesting source emitting 511 keV line predominantly, for example, a locally dense region aligned with stellar streams, Galactic winds, or transient events such as binary neutron star mergers \citep{Fuller2019} and calcium-rich supernovae \citep{Perets2014}, or nearby bubbles \citep{Siegert2024_LocalBubble}.

Although it is statistically significant, we need to be careful about artifacts of the image deconvolution. 
As seen at the bottom of Fig.~\ref{fig_bestimage_w_max1e-3_annotation}, the exposure is not uniform in high-latitude regions.
Furthermore, when splitting the full dataset into four independent periods, we found that this structure appears obviously in the image of the last period (years 2017--2023, see Appendix~\ref{sec_cumulative} and Fig.~\ref{fig_split_analysis}).
From January 2022 to October 2022, SPI monitored the blazar Mkn 501, located at ($l$, $b$) = (63.6$^{\circ}$, 38.9$^{\circ}$) and apart by $\sim 6^\circ$ from the center of the blob-like structure.
This feature may arise from systematic uncertainties in the SPI response function at the boundaries of its field of view rather than being a real astronomical source. 
Additionally, we observed a similar structure when performing image reconstruction in an adjacent energy band (526--532 keV) with a flux of $\sim 4 \times 10^{-5}$ ph cm$^{-2}$ s$^{-1}$.

At ($l$, $b$) = ($260$, $-60$) degrees, another structure can be seen, annotated as ``South Blob'' in Fig.~\ref{fig_bestimage_w_max1e-3_annotation}. 
However, the non-uniform exposure could also affect this structure.
To conclude the existence of these structures, uniform exposure is crucial to minimize such systematic uncertainties in image reconstruction for future 511 keV observations.
In any case, this southern emission feature is too far away from the Magellanic Clouds. Interestingly, \citet{Siegert2020} saw a similar emission feature at high negative latitudes with the COSI balloon, which they also attributed to image deconvolution artifacts.

Other emission features in Fig.~\ref{fig_bestimage_w_max1e-3_annotation}, such as around the Cepheus region ($l \sim 120^\circ$) or at the position of M82 (in particular deep exposure of the type Ia supernova SN2014J, e.g, \cite{Diehl2014_SN2014J}; $(l,b) \sim (140^\circ,40^\circ)$) appear as reasonable positron annihilation regions, but are not significant.
Local features, such as Loop I or the North Polar Spur, are also not aligned with any of these serendipitous emission features.
Strong individual hotspots from the filamentary structure of the Local Bubble \citep{Siegert2024_LocalBubble} may explain some residual emission, though not at the flux levels quoted here.

\begin{table*}[htbp]
    \centering
    \caption{Measured fluxes and significance of the 511 keV line emission from individual regions.}
    \begin{tabular}{c|cc|cc|c}
        \hline
        Name & \multicolumn{2}{c|}{Region} & 99\% Flux Upper Limit & Best-Fit Flux & Chance Probability (\%) \\
        & $l$ (deg.) & $b$ (deg.) & \multicolumn{2}{c|}{($\times 10^{-5}$ ph cm$^{-2}$ s$^{-1}$)} & \\
        \hline
        Cygnus & [$60$, $80$] & [$-15$, $15$]          & $<6.5$ & $4.2 \pm_{\mathrm{stat}} 2.0 \pm_{\mathrm{sys}} 2.0$ & $6.4 \times 10^{-2}$ \\
        Upper Scorpius & [$340$, $360$] & [$25$, $40$] & $<2.6$ & $1.4 \pm_{\mathrm{stat}} 0.8 \pm_{\mathrm{sys}} 0.5$ & $8.0 \times 10^{-2}$ \\
        Orion & [$190$, $220$] & [$-50$, $-10$]        & $<7.0$ & $2.7 \pm_{\mathrm{stat}} 1.3 \pm_{\mathrm{sys}} 0.8$ & $4.6 \times 10^{-1}$ \\
        Orion + Perseus & [$150$, $220$] & [$-50$, $-10$] & $<12.1$ & $8.6 \pm_{\mathrm{stat}} 2.4 \pm_{\mathrm{sys}} 2.5$ & $1.1 \times 10^{-1}$ \\
        Vela & [$260$, $290$] & [$-15$, $15$]          & $<8.3$ & $4.4 \pm_{\mathrm{stat}} 2.1 \pm_{\mathrm{sys}} 3.1$ & $1.9 \times 10^{-1}$ \\
        \hline
    \end{tabular}
    \tablefoot{The chance probabilities are derived from background-only bootstrap samples.}
    \label{tab_flux_indivisuals}
\end{table*}

\section{Discussion}
\label{sec_discussion}

\subsection{Comparison with previous studies}

Our reconstructed 511 keV map confirms the basic morphological features reported in previous studies: a bright bulge emission extending to about 10 degrees and a fainter disk emission along the Galactic plane. 
The measured fluxes from the bulge and disk regions show good agreement with recent measurements by \cite{Siegert2016} and \cite{Skinner2015}, demonstrating the robustness of our image deconvolution.

While \cite{Siegert2016} and \cite{Skinner2015} mainly employed model-fitting approaches with predefined spatial templates, the image deconvolution approach adopted in this work allows us to investigate the spatial distribution without assuming specific functional forms.
This is particularly important because, although the model fitting approach provides precise flux measurements under assumed morphologies, it can introduce biases and miss local structures due to constraints about the spatial structure through assumed models.

\begin{figure}
    \centering
    \includegraphics[width=0.9\linewidth]{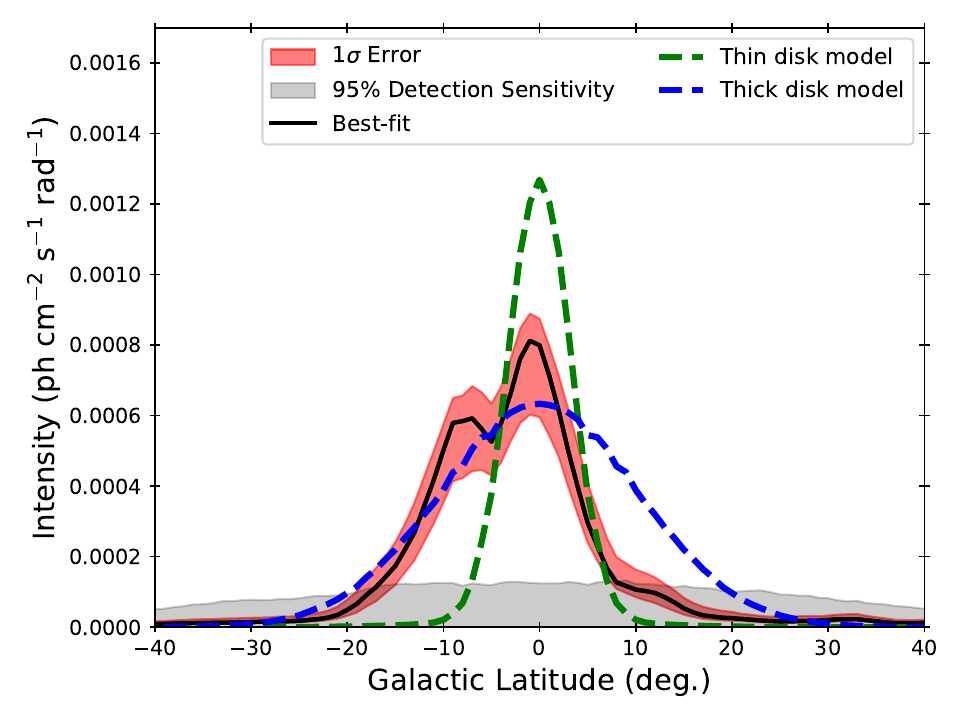}
    \caption{Latitude profile of the 511 keV line emission integrated over the disk regions ($l = 20 -50, 310-340$ degrees). For comparison,
    the latitude profiles of the thin (green) and thick (blue) disk models are also shown.}
    \label{fig_integrated_flux_comp_disk}
\end{figure}

The latitude profile of the disk emission can illustrate the advantage of our model-independent approach.
Figure~\ref{fig_integrated_flux_comp_disk} shows the latitude distribution of the 511 keV emissions integrated over the disk regions ($l = 20^{\circ}-50^{\circ}$ and $310^{\circ}-340^{\circ}$, corresponding to the eastern and western disk regions in Table~\ref{tab_global_flux}).
Previous model-fitting studies mainly characterized the disk morphology using two-dimensional Gaussian profiles with scale heights ($\sigma$ parameter for the vertical extension) of 3$^{\circ}$ for the thin disk model or 10.5$^{\circ}$ for the thick disk model.
However, our reconstructed profile may suggest an asymmetric structure; the 511 keV emission extends further at the negative latitude than at the positive latitude.
Such a potential asymmetric vertical structure of the disk could provide new insights into positron propagation or source distribution in the Galactic disk.
If real, the extended structure in the negative latitude region should be (partially) generated by sources preferentially distributed in that region.

\begin{figure}
    \centering
    \includegraphics[width=0.9\linewidth]{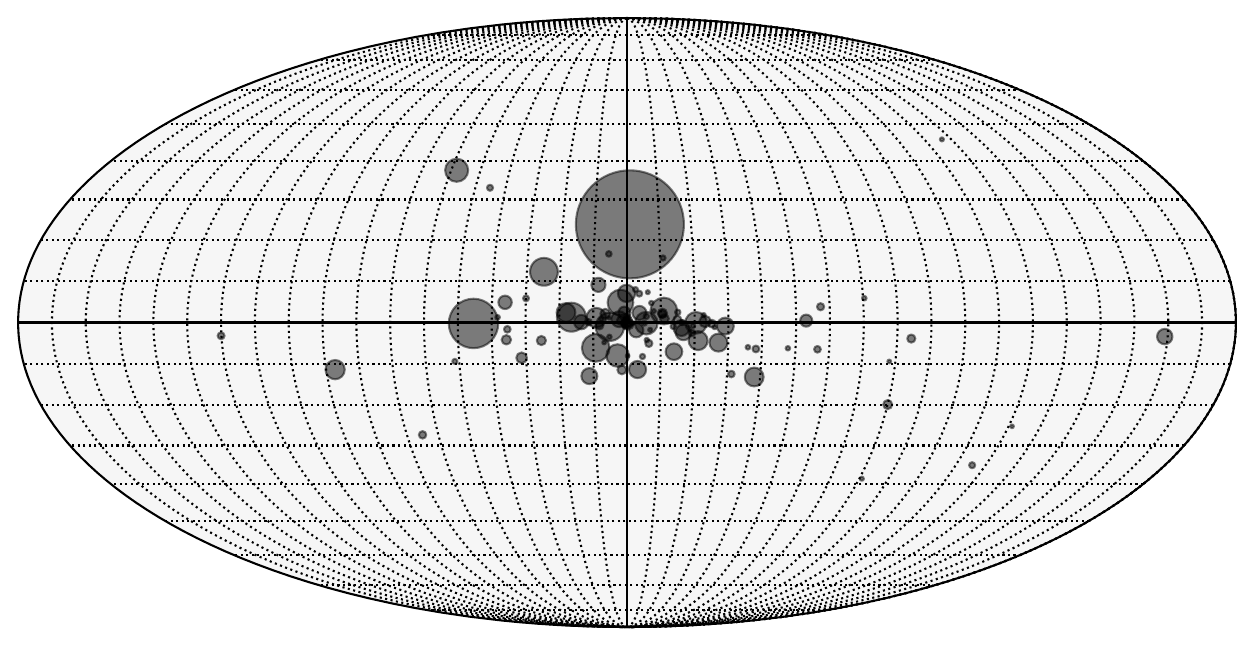}
    \caption{Distribution of the LMXBs in the Swift BAT 157-month survey catalog. The radius of each circle is proportional to the X-ray flux of its corresponding source.}
    \label{fig_LMXB_distribution}
\end{figure}

For example, \cite{Weidenspointner2008} suggested that low-mass X-ray binaries with hard X-ray emissions (hard LMXBs) explain the 511 keV emission in the disk region.
They reported a flux ratio of $\sim2$ between negative and positive longitude regions, and hard LMXBs show a similar asymmetric distribution along the Galactic plane.
To revisit this hypothesis using our 20-year dataset, we examined the spatial distribution of LMXBs using the Swift-BAT 157-month catalog \citep{lien2025}. 
The catalog shows 71 LMXBs in the eastern hemisphere ($l < 0^\circ$) versus 46 in the west ($l > 0^\circ$), confirming an asymmetric distribution similar to previous findings (see Fig.~\ref{fig_LMXB_distribution}). 
However, the interpretation depends critically on the assumed positron production efficiency. 
When considering disk sources within $|b| < 20^\circ$,
weighting by X-ray flux gives an east/west ratio of $\sim0.6$, which is opposite to earlier claims.
On the other hand, if we assume that the positron annihilation rate is proportional to $L_\mathrm{X}^{0.5}$, where $L_\mathrm{X}$ is the X-ray luminosity, the east/west ratio becomes $\sim1.4$.
Here, we extracted sources whose distances were estimated in \cite{Avakyan2023}.
This demonstrates that understanding the luminosity dependence of positron production is crucial for interpreting spatial asymmetries. 
Regarding the latitude profile, LMXBs in the eastern disk region ($l = -50^\circ$ to $-20^\circ$) are indeed concentrated at negative latitudes, qualitatively consistent with the southern offset in our 511 keV map.
We note that, if 511 keV emission simply scales with X-ray luminosity regardless of individual LMXB properties, Sco X-1 (the brightest LMXB at $b \approx 24^\circ$) would produce stronger emission.
These results indicate that drawing robust conclusions about LMXB contributions requires detailed modeling of positron production mechanisms in jets, escape processes from binary systems, and propagation through different ISM environments for each source, which simple estimations above cannot capture.

\subsection{On the hints of 511 keV emission from massive stars}

Our image reconstruction reveals several intriguing features in the 511 keV emission map.
In the bulge region, we find an asymmetric structure. The emission may extend further in the positive latitude region while showing a sharper drop-off in the negative latitude direction.
It is accompanied by the possible chimney-like structure around ($l$, $b$) = (350$^{\circ}$, 30$^{\circ}$), spatially coincident with the Upper Scorpius region.
A similar asymmetry was also reported in the 511 keV mapping with OSSE/CGRO observations, referred to as the ``OSSE 511\,keV fountain'' \citep{Purcell1997}.
This suggests that the positron production or annihilation is enhanced locally in this region.

A possible interpretation is that 511 keV originates from positrons produced via the $\beta^{+}$ decay from $^{26}$Al in the Scorpius-Centaurus association.
A quantitative comparison with $^{26}$Al supports this scenario.
By comparing the 511 keV fluxes between a northeastern region ($l = $ [340$^{\circ}$, 360$^{\circ}$], $b = $ [10$^{\circ}$, 40$^{\circ}$]) and southeastern region ($l = $ [340$^{\circ}$, 360$^{\circ}$], $b = $ [-40$^{\circ}$, -10$^{\circ}$]),
we estimated the excess emission from the Scorpius-Centaurus association to be $\approx 4 \times 10^{-5}$ ph cm$^{-2}$ s$^{-1}$.
This is comparable to the $^{26}$Al flux of $6-8 \times 10^{-5}$ ph cm$^{-2}$ s$^{-1}$ reported from \cite{Diehl2010,Krause2018} at more than 5$\sigma$ significance.
Theoretically, the fraction of the 511 keV flux to the $^{26}$Al flux is $\sim$0.4, derived from the branching ratio of $\beta^{+}$ decay of $^{26}$Al (0.82), the positronium fraction (0.9-1.0), and two-photon emission from a positronium.
Considering non-negligible errors on the fluxes, the derived 511 keV flux is consistent with the scenario that $^{26}$Al in the Scorpius-Centaurus association produces the locally enhanced 511 keV emission.

This possibility can provide a unique opportunity to study positron propagation in the local interstellar medium (ISM). 
A positron produced from $\beta^{+}$ decay of $^{26}$Al has a mean kinetic energy of 0.543 MeV with an end-point energy of 1.173 MeV \citep{Al26data}.
When such low-energy positrons are injected into the ISM, they lose energy through ionization losses or undergo in-flight annihilation mainly through positronium formation via charge exchange with neutral atoms or molecules \citep{Guessoum2005,Prantzos2011}.
The latter is the dominant annihilation process for positrons in neutral media.
In either case, positrons must cool down to $\sim$100 eV before in-flight positronium formation or thermalization can begin, and this cooling process requires $O$(0.1-10) Myr depending on the ISM conditions  \citep[e.g.,][]{Panther2018}.
If the positrons are not annihilated in-flight, they thermalize with the surrounding gases and then form positronium mainly through radiative recombination with ambient electrons or charge exchange with neutral atoms, before annihilating.
This thermalization and positronium formation process requires additional time, typically on the order of the cooling process.
Given that the distance of the Scorpius-Centaurus association is about 100 pc \citep{deGeus1992,deZeeuw1999,Preibisch1999,Gaia2020,Grandjean2023} and the observed 511 keV emission is confined to this region with an extension of $\lesssim 10$ degrees in the reconstructed image, corresponding to $\sim 20$ pc,
we can constrain the diffusion coefficient of the low-energy positrons.
Considering that the mean square displacement on a projected two-dimensional plane is given by $4 D t $, where $D$ is the diffusion coefficient and $t$ is the time of the diffusion \citep{Landau1987Fluid}, 
$D$ is bounded by
\begin{eqnarray}
   D \lesssim O(10^{26})~\mathrm{cm^{2}~s^{-1}}~.
\end{eqnarray}
The propagation of the low-energy positron is complex and remains poorly understood \citep{Jean2009}.
If future observations confirm this scenario, it would provide a new way to measure the low-energy cosmic-ray propagation in the ISM.
Also, such a relatively small diffusion coefficient broadly impacts the understanding of the morphology of the 511 keV emission. 
For instance, \cite{Pedro2024} modeled the 511 keV emission from a MeV-scale dark matter model and showed that the diffusion coefficient of the positrons could change the resulting 511 keV emission map significantly.

When investigating the potential 511 keV emission from massive stars with future observations, we note that a critical challenge will be distinguishing local excesses from global disk emissions.
As shown in Table~\ref{tab_global_disk_contamination}, the expected contributions from the thin and thick disk models vary significantly by region of interest.
For regions near the Galactic plane and center, such as Cygnus and Vela, the disk contribution is substantial ($\sim 10^{-4}$ ph cm$^{-2}$ s$^{-1}$), making it challenging to identify any additional local component.
We need to understand the global structure of the 511 keV disk emission well enough to separate the local components from the global disk component.
On the other hand, regions at higher latitudes such as Orion and particularly Upper Scorpius would be ideal regions to avoid this issue since the contamination from the global component is much lower ($\lesssim 10^{-6}$ ph cm$^{-2}$ s$^{-1}$).
Thus, the Scorpius-Centaurus association, where hints of 511 keV emission are found, would be a particularly important target for future studies to identify the origin of the 511 keV emission and investigate positron propagation.

While we have focused on massive stars as a potential source to explain the possible asymmetric bulge structure, we note that other scenarios could also explain it.
For example, the chimney-like structure in the Upper Scorpius region spatially 
coincides with the inner regions of the Fermi and eROSITA bubbles \citep{Su2010,Fermi2014,Predehl2020}, suggesting possible connections to past energetic outbursts from the Galactic center \citep[e.g.,][]{Totani2006,Crocker2011}. 
However, the inferred north-south asymmetry may not favor this 
scenario, since these bubble structures are symmetric about the Galactic plane. More sensitive observations, such as with COSI, are needed to test this possibility conclusively.
Additionally, the asymmetry near the Galactic center may be explained by positron production through cosmic-ray interactions with asteroids \citep{Siegert2024}.
Regarding the Upper Scorpius emission, we observe hints of this emission in two different epochs (see Fig.~\ref{fig_split_analysis}A), which could support either steady-state or variable emission from this region, depending on the exact exposure configuration given the variable sources within the Solar System.
More detailed temporal analysis will be presented in a future work.

\begin{table}[htbp]
    \centering
    \caption{Expected flux contributions from the thick and thin disk models in the individual source regions.}
    \begin{tabular}{c|cc}
        \hline
        Name & Thick Disk & Thin Disk \\
        & \multicolumn{2}{c}{(ph cm$^{-2}$ s$^{-1}$)} \\
        \hline
        Cygnus & $ 9.3 \times 10^{-5}$ & $ 9.2 \times 10^{-5}$\\
        Upper Scorpius & $ 2.4\times 10^{-6}$ & $ 7.3 \times 10^{-7}$\\
        Orion & $ 2.0 \times 10^{-6}$ & $ 2.0 \times 10^{-8}$\\
        Orion + Crab & $ 2.3 \times 10^{-6}$ & $ 2.4 \times 10^{-8}$\\
        Vela & $ 1.0 \times 10^{-4}$ & $ 1.2\times 10^{-4}$\\
        \hline
    \end{tabular}
    \label{tab_global_disk_contamination}
\end{table}

\subsection{Implications for the future MeV gamma-ray observations}

Our analysis has highlighted two key challenges in studying the 511 keV emission with INTEGRAL/SPI and interpreting the results. First, the non-uniform exposure across the sky prevents us from concluding whether some faint structures in the reconstructed image are real. As shown in Fig.~\ref{fig_exp_map}, the exposure time varies considerably, with the maximum difference in the exposure map being about three orders of magnitude. This non-uniformity can cause systematic uncertainties in the image deconvolution, making it difficult to confirm potential features. This issue is particularly problematic for the high-latitude structures we detected around ($l$, $b$) = (70$^{\circ}$, 40$^{\circ}$) and (260$^{\circ}$, -60$^{\circ}$), whose reliability remains uncertain.

The second challenge is confirming the marginal detections reported in this work, especially those from individual regions. Our analysis revealed several interesting features at the 2$\sigma$ level, such as the possible 511 keV emission from the Cygnus and Upper Scorpius regions. While these results are tantalizing and could provide new insights into positron production and propagation, their relatively low significance requires further observations for confirmation.

We need uniform all-sky observations with improved sensitivity to overcome these limitations in the current observations. The upcoming MeV gamma-ray satellite COSI \citep{COSIofficial} can address these issues. COSI is a NASA Small Explorer satellite mission planned for launch in 2027. It is designed to survey the entire sky in the energy range of 0.2--5 MeV using a wide field-of-view Compton telescope, which can instantaneously cover 25\% of the sky. With this observational mode, the exposure map is expected to be more uniform than that of SPI, which would help confirm the high-latitude structures reported in our analysis. In particular, the improved spatially uniform observations would be vital to verify the possible correlation between 511 keV and $^{26}$Al emission in the Upper Scorpius region, which could provide observational constraints on the low-energy positron propagation in the ISM.

With two years of observations, COSI will achieve a line sensitivity of $\sim 10^{-5}$ ph cm$^{-2}$ s$^{-1}$ at 511 keV. This sensitivity would exceed that of INTEGRAL/SPI across most of the sky.
Such improved sensitivity would be sufficient to investigate the features we report at marginal significance.

\section{Conclusions}
\label{sec_conclusion}

We have performed image reconstruction analysis of the 511 keV line emission using 20 years of INTEGRAL/SPI data with the current highest available photon statistics.
Our analysis employed the Richardson-Lucy algorithm for image deconvolution and implemented the bootstrap method to evaluate statistical uncertainties.
Additionally, we evaluated systematic uncertainties in the choice of parameters for the image deconvolution algorithm.
Such analysis required considerable computational resources, which we addressed through GPU acceleration.

The reconstructed image confirms the basic morphological features reported in previous model-fitting studies: a bright central component within $\sim$ 3 degrees from the Galactic center, a broad bulge component, and an elongated disk component along the Galactic plane.
These results demonstrate the robustness of our image deconvolution approach.

Our analysis also suggests intriguing hints of new spatial features in the 511 keV emission.
Notably, we find that the bulge emission may extend asymmetrically toward the northeastern direction with a faint blob.
This extension is spatially coincident with the region where the enhanced $^{26}$Al emission was reported from the Scorpius–Centaurus association.
This correlation could indicate a first detection of 511 keV emissions associated with $^{26}$Al from massive stars, although its significance remains marginal.

Detailed interpretation of the spatial structures is limited by two factors in the current observations.
First, the non-uniform exposure across the sky may cause artifacts, such as blobs in high-latitude regions.
Second, the sensitivity of current observations only allows us to detect spatial features at a marginal significance beyond the bulge and disk components.
Future MeV gamma-ray observations, such as with COSI, will address these issues with improved sensitivity and uniform exposure through all-sky survey capabilities.
Such observations will enable us to confirm the reported tantalizing spatial structures and reveal the origin of the positron production and its propagation in our Galaxy.

\begin{acknowledgements}
The authors would like to thank Israel Martinez-Castellanos and Aravind Valluvan for their effort on the development of the image reconstruction framework in the cosipy library, and Yoshiyuki Inoue for valuable comments on the manuscript of this work.
HY and SM acknowledge support by the Bundesministerium f\"{u}r Wirtschaft und Klimaschutz via the Deutsches Zentrum f\u{u}r Luft- und Raumfahrt (DLR) under contract number 50 OO 2219.
HY is supported by JSPS KAKENHI Grant Number 23K13136.
\end{acknowledgements}

\bibliographystyle{aa} 

\begin{appendix}

\section{Data selection criteria}
\label{sec_data_selection}

The data selection is based on several criteria. Here, we use absolute and relative count rates from the INTEGRAL Radiation Environment Monitor (IREM), in particular the TC3 and S32 rate \citep{Hajdas2003}, which turned out to filter several solar flares and solar particle events. In addition, we use the ratio between the ACS rate versus the total number of Ge detector events registered, as well as the total Ge detector events versus the events that only hit a single detector. The selection of temperature is based on the difference between the cooling plates, not the absolute temperature. This filters anomalies after the annealing phases. The selection of high voltage strictly follows the mean voltage for each camera configuration and the PI settings. For each of these parameter values as a function of time, we perform a median filter clipping. Here, the window of the median filter comprises 5000 pointings, with varying thresholds in units of the standard deviation of the parameters. The thresholds are described in Table~\ref{tab_selection_threshold}.
Typically, the temperature difference of the SPI cooling plates is less than 0.1 K, and the running mean of the high voltage is less than 0.05 V.
Finally, we deselect entire revolutions in which either strong transients occurred, which cannot be properly modeled by our background method, and the revolutions in which the detector failures happened. This includes revolutions 140, 214, 215, 775, 776, 930, 931, 236, 354, 711, 1147, 1149, 1604, 1554, 1555, 1556, 1557, 1558, 2682, 2684, 2685. 

\begin{table}[htbp]
    \centering
    \caption{Thresholds in the median filter clipping of the data selection.}
    \begin{tabular}{c|c}
    \hline
    Parameters & threshold [$\sigma$] \\
    \hline
    TempColdPlt1 - TempColdPlt3  & 2 \\
    ACSRate/GeSatTot & 0.5 \\
    IREM\_TC3\_Rate & 0.75 \\
    IREM\_S32\_Rate & 0.75 \\
    HVDetMean & 0.1 \\
    \hline
    \end{tabular}
    \label{tab_selection_threshold}
\end{table}

\section{Details of the parameter optimization of the image deconvolution algorithm}
\label{sec_details_parameter_optimization}

\begin{figure}
    \centering
    \includegraphics[width=0.95\linewidth]{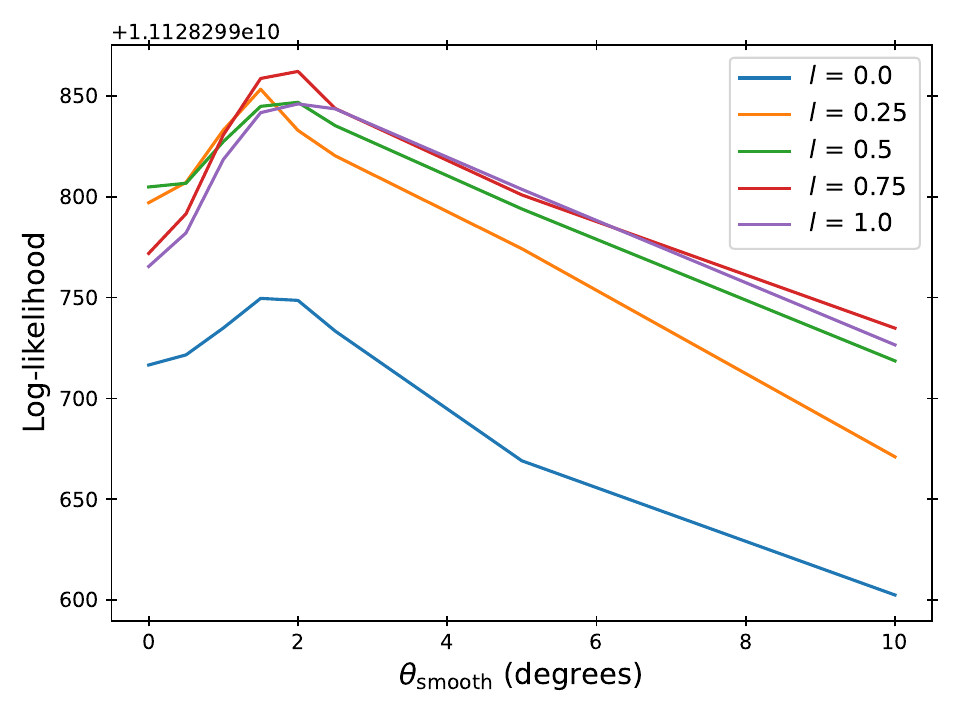}
    \caption{Log-likelihood curve as a function of the smoothing kernel size $\theta_{\mathrm{smooth}}$ for different weighting power $l$. The optimal parameters are found to be $l = 0.75$ and $\theta_{\mathrm{smooth}} = 2.0$ degrees, which gives the maximum in this curve.}
    \label{fig_loglikelihood}
\end{figure}

Figure~\ref{fig_loglikelihood} shows the log-likelihood as a function of $\theta_{\mathrm{smooth}}$ and $l$.
We can see a peak around $\theta_{\mathrm{smooth}} = 2$ degrees, which is close to SPI's angular resolution of 2.7 degrees.
The weighting power $l$ shows optimal values between 0.5 and 1.0, with the maximum likelihood achieved at $l = 0.75$.

\begin{figure}
    \centering
    \includegraphics[width=0.95\linewidth]{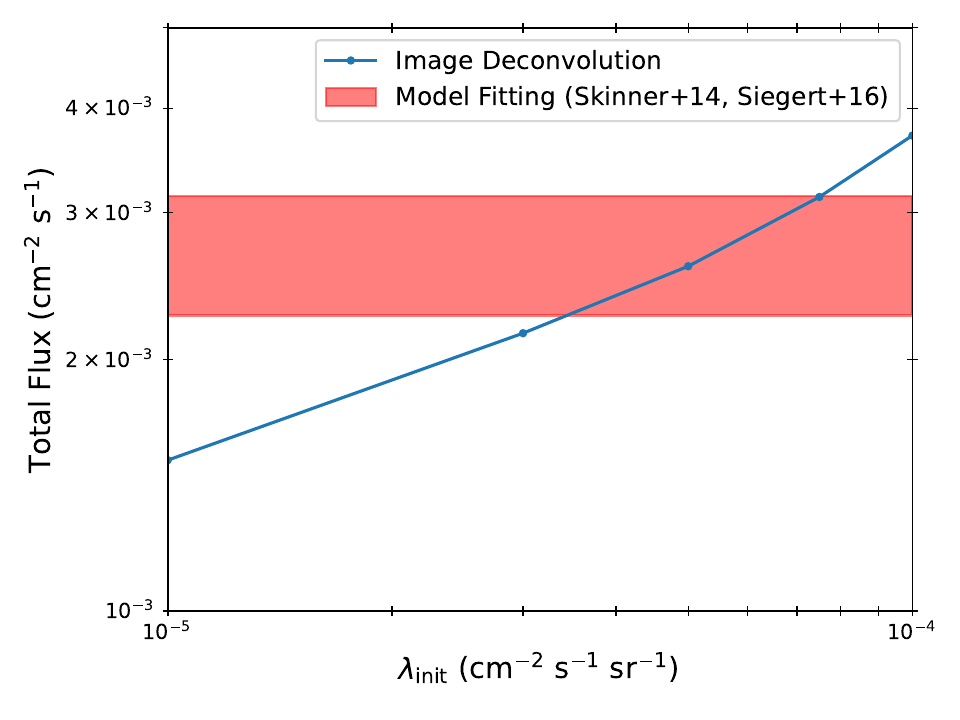}
    \caption{Total flux of the reconstructed image as a function of the initial intensity $\lambda_{\mathrm{init}}$. The horizontal band indicates the range of total flux values reported from previous model fitting studies \citep{Skinner2015,Siegert2016}.}
    \label{fig_initial_flux_dependence}
\end{figure}

We also investigated how the reconstructed image is affected by $\lambda_{\mathrm{init}}$.
The choice of $\lambda_{\mathrm{init}}$ significantly influences the results.
Figure~\ref{fig_initial_flux_dependence} shows how the total reconstructed flux, i.e., the flux integrated over the sky, varies with different values of the intensity in the initial map.
When starting with large values, the reconstructed image tends to show numerous artifacts as the algorithm attempts to suppress the excess initial flux in each pixel.
On the other hand, with very small values, the reconstruction converges very slowly. Consequently, reconstructed images emphasize only the bright structures and miss the fainter ones we are interested in.
Comparing with the previous model-fitting studies \citep{Skinner2015,Siegert2016}, we adopted $\lambda_{\mathrm{init}} = 5 \times 10^{-5}$ cm$^{-2}$ s$^{-1}$ sr$^{-1}$ as it yields a consistent total flux with the previous measurements.

\section{Stability of the image deconvolution depending on $\alpha_{\mathrm{max}}$}
\label{sec_dep_max_alpha}

\begin{figure}
    \centering
    \includegraphics[width=0.95\linewidth]{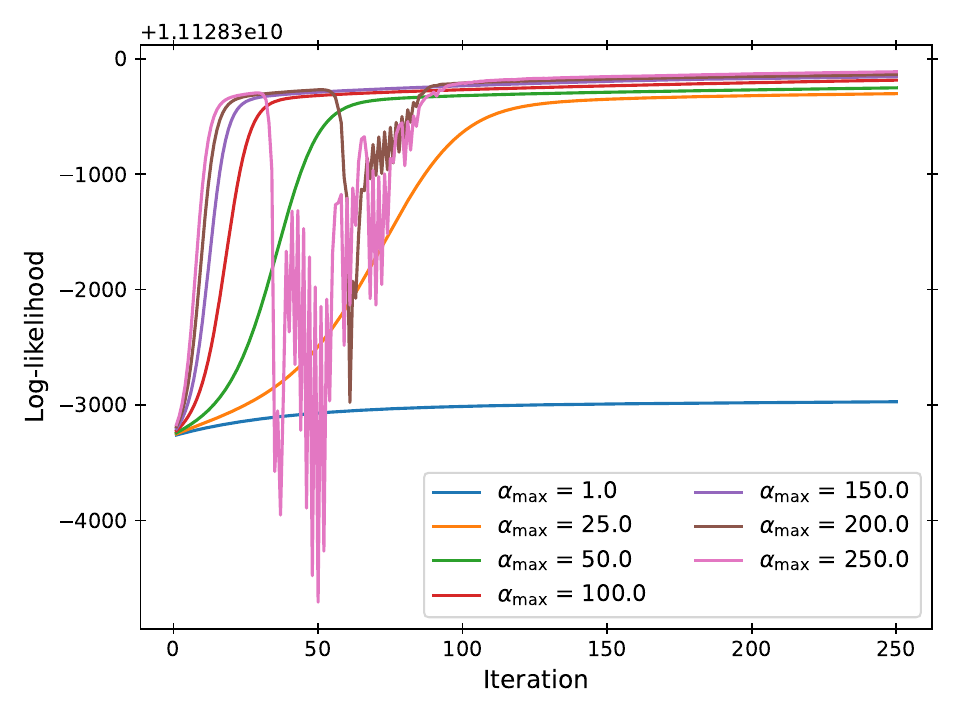}
    \caption{Log-likelihood curve as a function of iteration number for different values of the acceleration parameter $\alpha_{\mathrm{max}}$.}
    \label{fig_dep_max_alpha}
\end{figure}

We investigate the stability of the image reconstruction depending on the maximum value of the acceleration parameter $\alpha_{\mathrm{max}}$ in the modified Richardson-Lucy algorithm.
Figure~\ref{fig_dep_max_alpha} shows how the log-likelihood evolves with iteration number for different values of $\alpha_{\mathrm{max}}$.
When $\alpha_{\mathrm{max}}$ is too large ($\geq 200$), the log-likelihood curve shows oscillation, indicating instability in the analysis while it converges eventually.
On the other hand, with small $\alpha_{\mathrm{max}}$ values, the convergence becomes slow.
Based on this analysis, we adopted $\alpha_{\mathrm{max}} = 100$ as the optimal value in our analysis, considering both stability and acceptable convergence speed.

\section{Comparison between the data and the reconstructed model}

\begin{figure}
    \centering
    \includegraphics[width=0.95\linewidth]{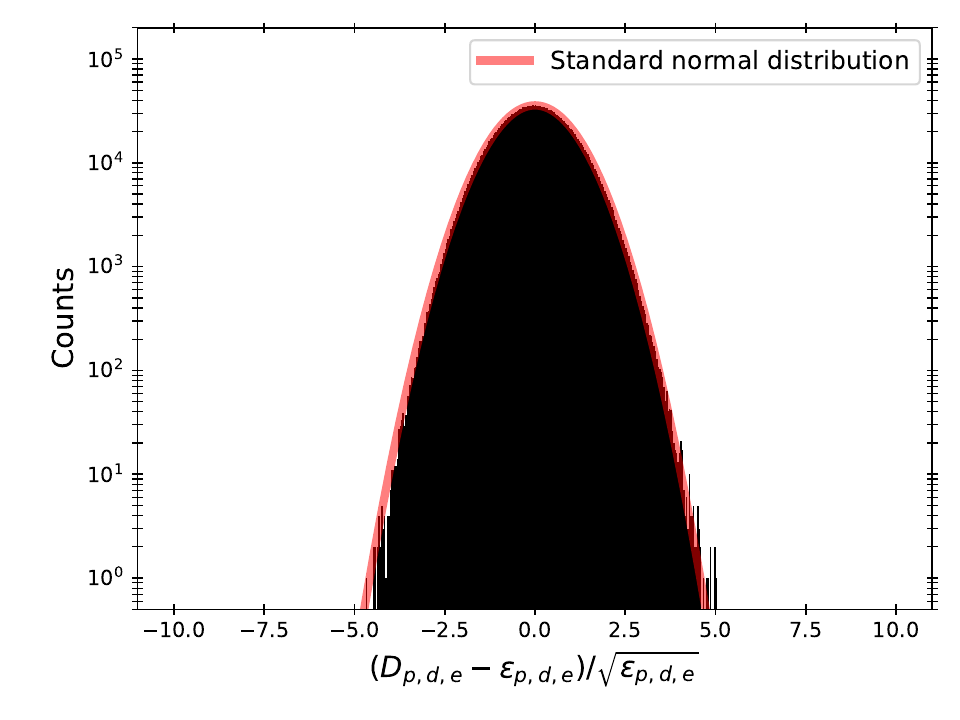}
    \caption{Distribution of normalized residuals $(D_{p,d,e} - \epsilon_{p,d,e}) / \sqrt{\epsilon_{p,d,e}}$, where $D_{p,d,e}$ is the observed count and $\epsilon_{p,d,e}$ is the expected count from the reconstructed image and background normalization. The solid red line shows a standard normal distribution for comparison.}
    \label{fig_residuals}
\end{figure}

We examine the reliability of our image reconstruction by comparing the observed data with the model predictions.
Figure~\ref{fig_residuals} shows the distribution of normalized residuals, the difference between detected and expected counts in each bin, divided by the square root of the expected count.
This distribution follows a standard normal distribution well, indicating that our model predicts the data well within statistical uncertainties.
The goodness of fit is quantified by $\chi^2 = 2241364$ with 2212539 degrees of freedom, yielding a reduced $\chi^2$ of 1.013, which is close to unity.

\begin{figure}
    \centering
    \includegraphics[width=0.95\linewidth]{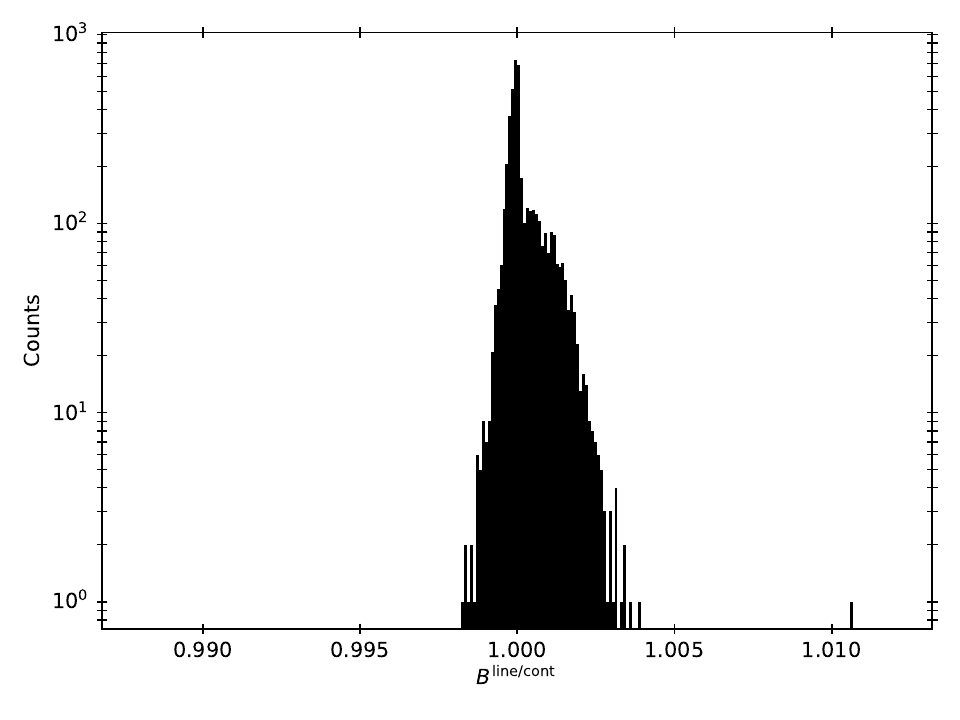}
    \caption{Distribution of the optimized background normalization factors ($B^{\mathrm{line}}$ and $B^{\mathrm{cont}}$).}
    \label{fig_background_normalization_dist}
\end{figure}

\begin{figure}
    \centering
    \includegraphics[width=0.95\linewidth]{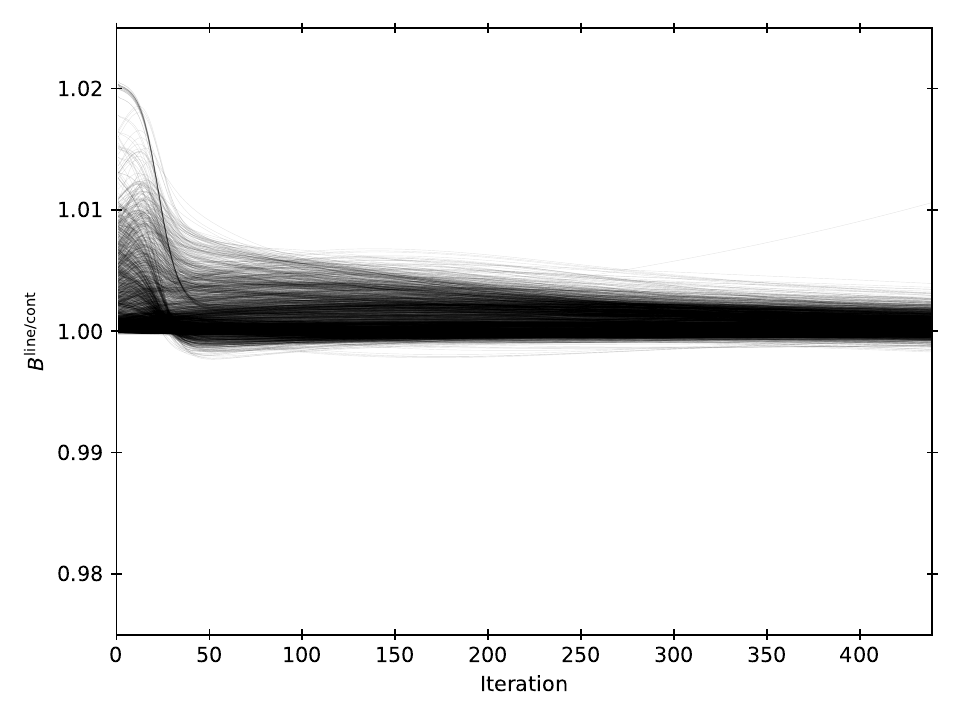}
    \caption{Evolution of the background normalization factors along the iteration number.}
    \label{fig_background_normalization_curve}
\end{figure}

The background normalization factors are also investigated.
Figure~\ref{fig_background_normalization_dist} shows their distribution after the image deconvolution, which is well-centered around unity.
We note that one data point is present around 1.01, which appears to be an outlier.
This corresponds to a pointing with a particularly short exposure time, where its orbital revolution contains only this single pointing.
We also show the evolution of the background normalization factors with iteration number in Fig.~\ref{fig_background_normalization_curve}.
While the background normalization for this revolution may not be fully converged, we have confirmed that excluding it from the analysis did not affect our results.

\section{Longitude profile with different integration ranges}
\label{sec_longitude_profile_more}

\begin{figure}
    \centering
    \includegraphics[width=0.95\linewidth]{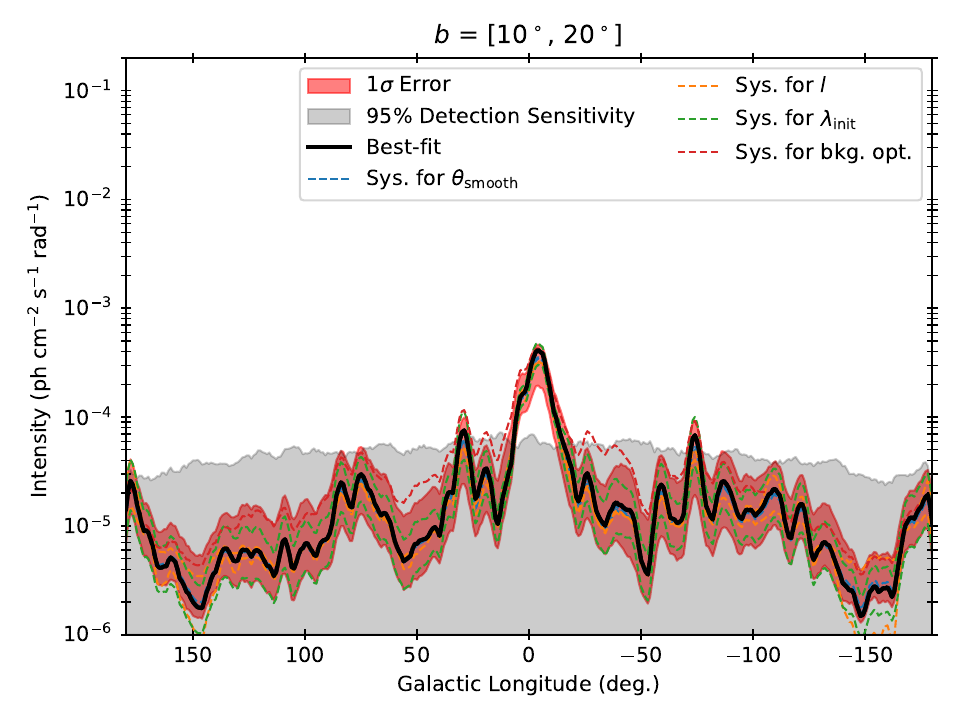}
    \includegraphics[width=0.95\linewidth]{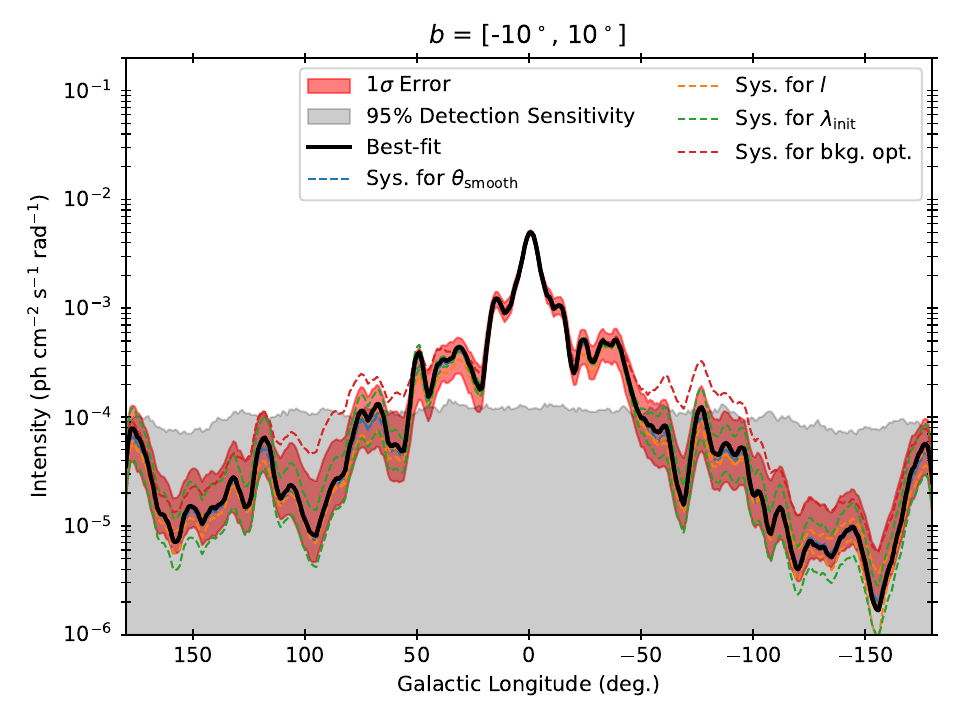}
    \includegraphics[width=0.95\linewidth]{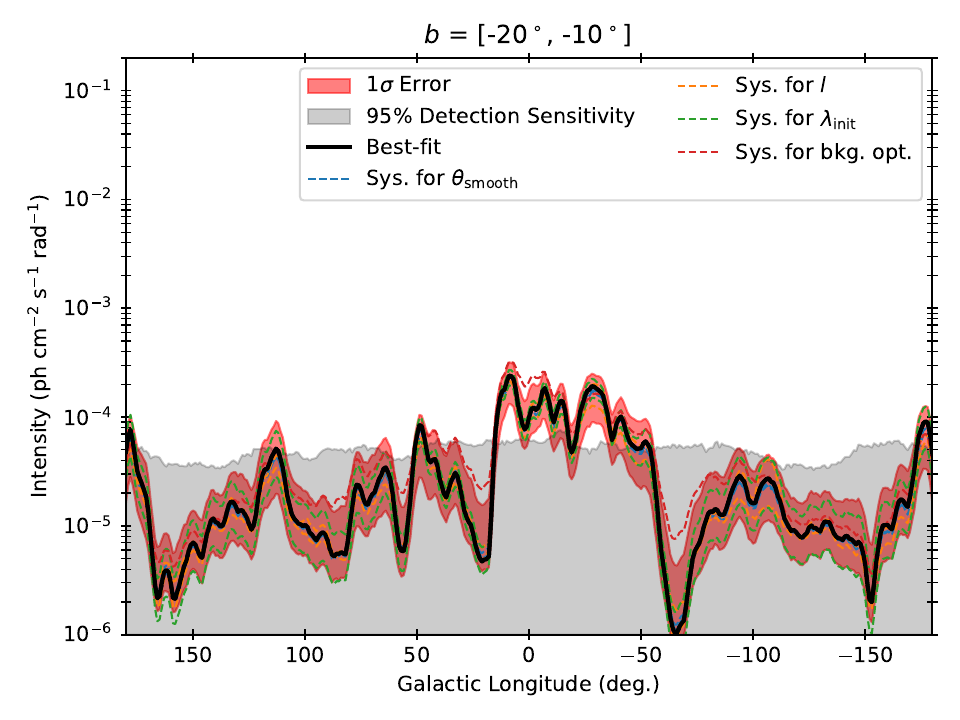}
    \caption{Same as Fig.~\ref{fig_profile_longitude}, but using different integration ranges; 
    from $b = 10$ to $20$ degrees (top), from $b = -10$ to $10$ degrees (middle), and from $b = -20$ to $-10$ degrees (bottom).}
    \label{fig_profile_latitude_w_different_range}
\end{figure}

We show the longitude profile with different integration ranges in Fig.~\ref{fig_profile_latitude_w_different_range}. While the profile from $b = -10^{\circ}$ to $10^{\circ}$ is similar to that shown in Fig.~\ref{fig_profile_latitude}, the longitude profile above $b = 10^{\circ}$ differs significantly from that below $b = -10^{\circ}$. Below $b = -10^{\circ}$, the profile near $l = 0^{\circ}$ is nearly flat, with intensity close to the sensitivity limit. In contrast, above $b = 10^{\circ}$, the profile exhibits a bright structure at $l < 0^{\circ}$, which peaks at $l \sim -4^{\circ}$. Together with the plateau observed at $b = 10^{\circ}$--$18^{\circ}$ in the latitude profile, this suggests flux enhancement in the northeastern region. This feature also hints at a chimney-like structure in the northern region. 

\section{Image analysis using different periods}
\label{sec_cumulative}

\begin{figure*}
    \centering
    \includegraphics[width=0.32\linewidth]{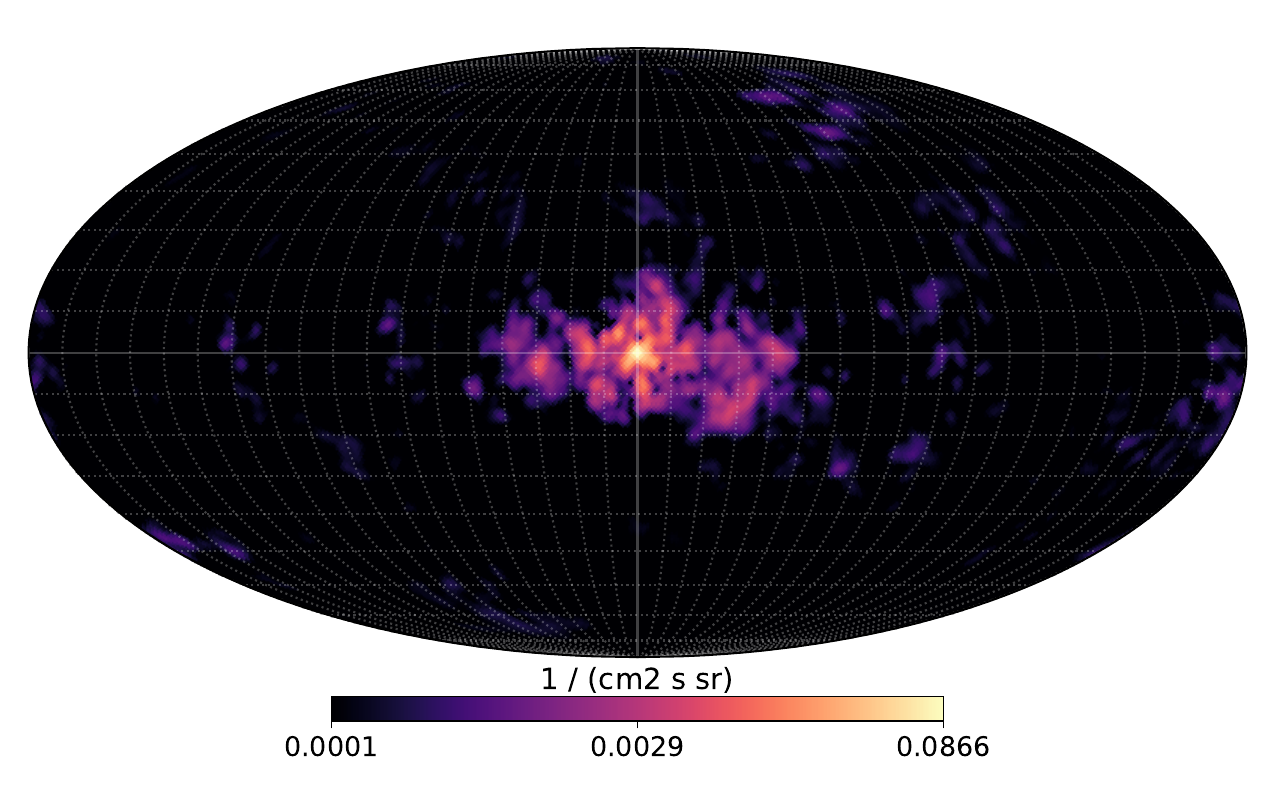}
    \includegraphics[width=0.32\linewidth]{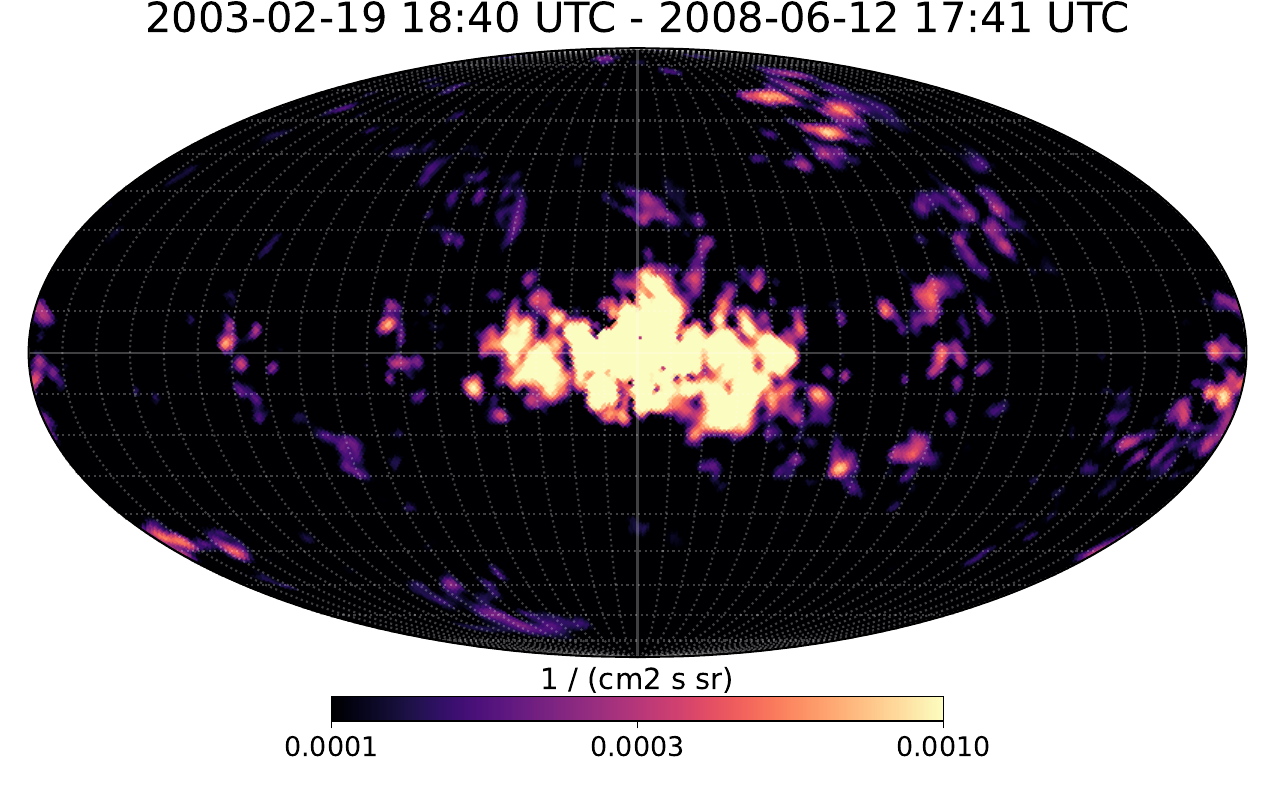}
    \includegraphics[width=0.32\linewidth]{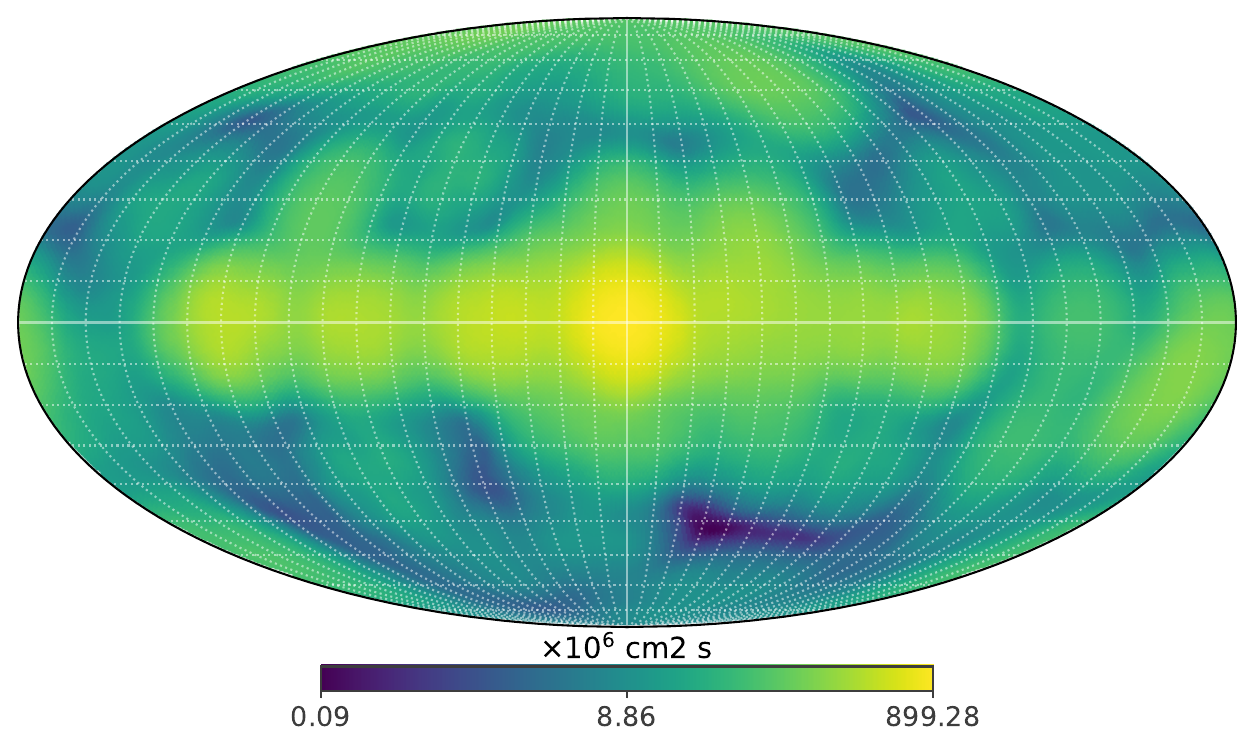}
    \includegraphics[width=0.32\linewidth]{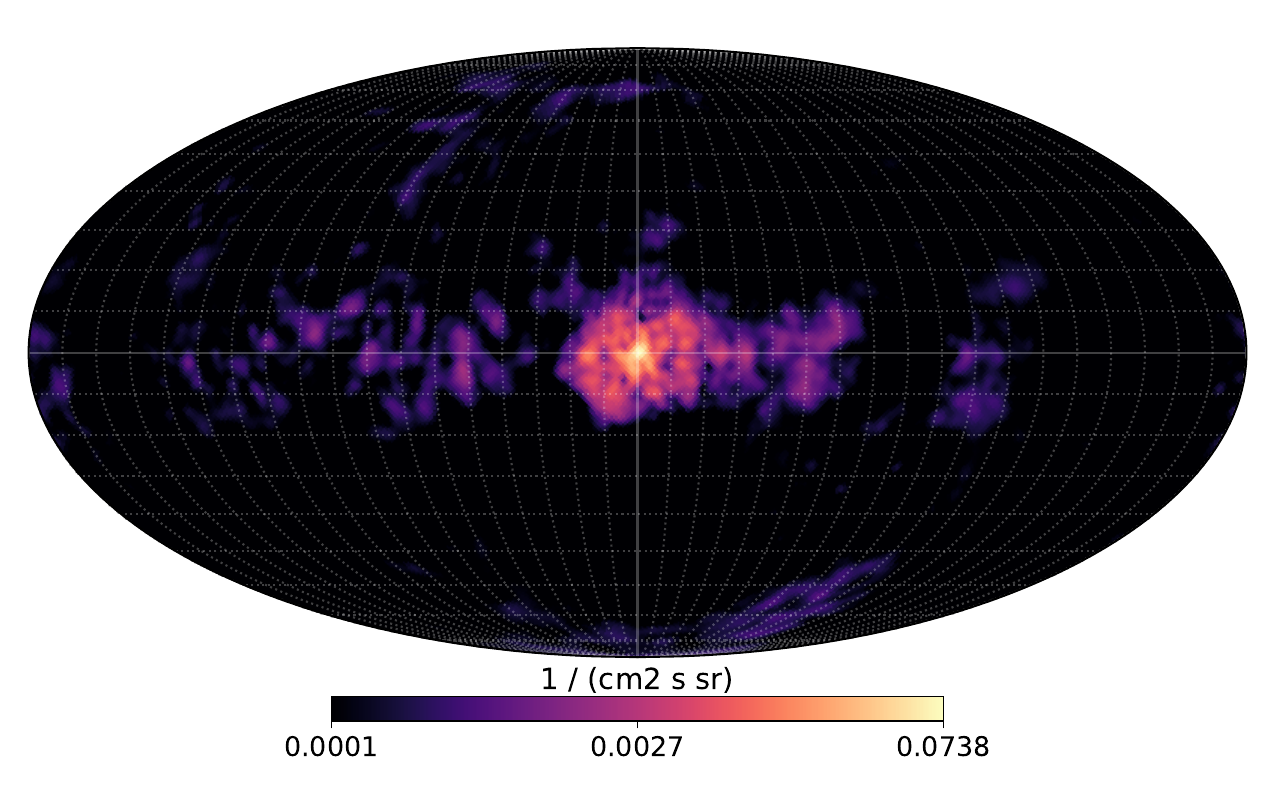}
    \includegraphics[width=0.32\linewidth]{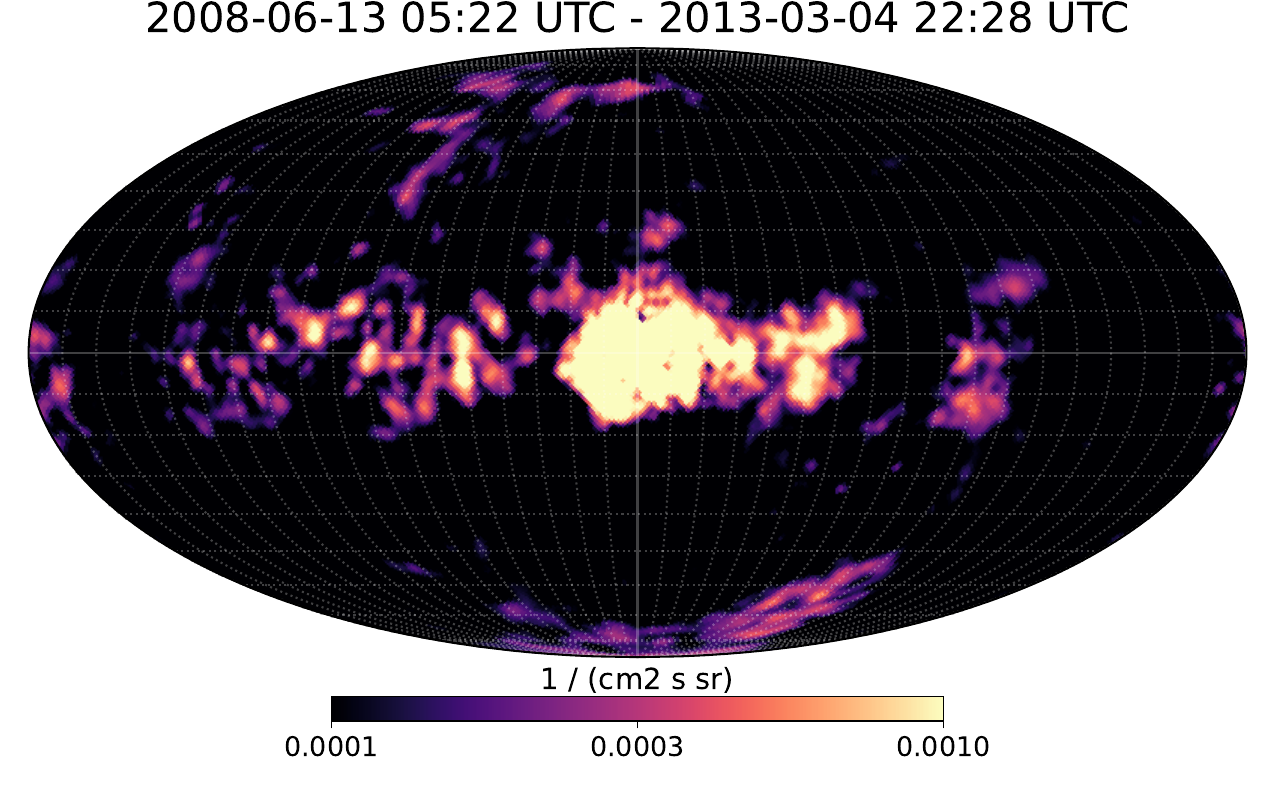}
    \includegraphics[width=0.32\linewidth]{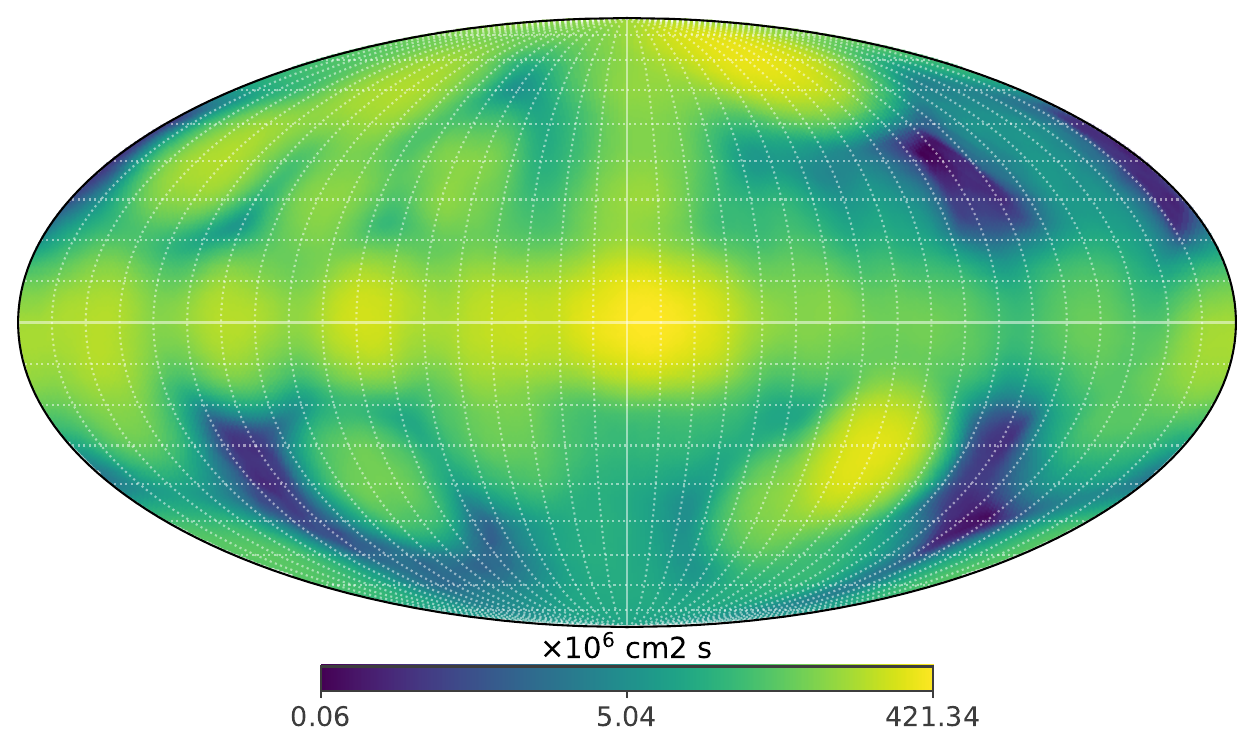}
    \includegraphics[width=0.32\linewidth]{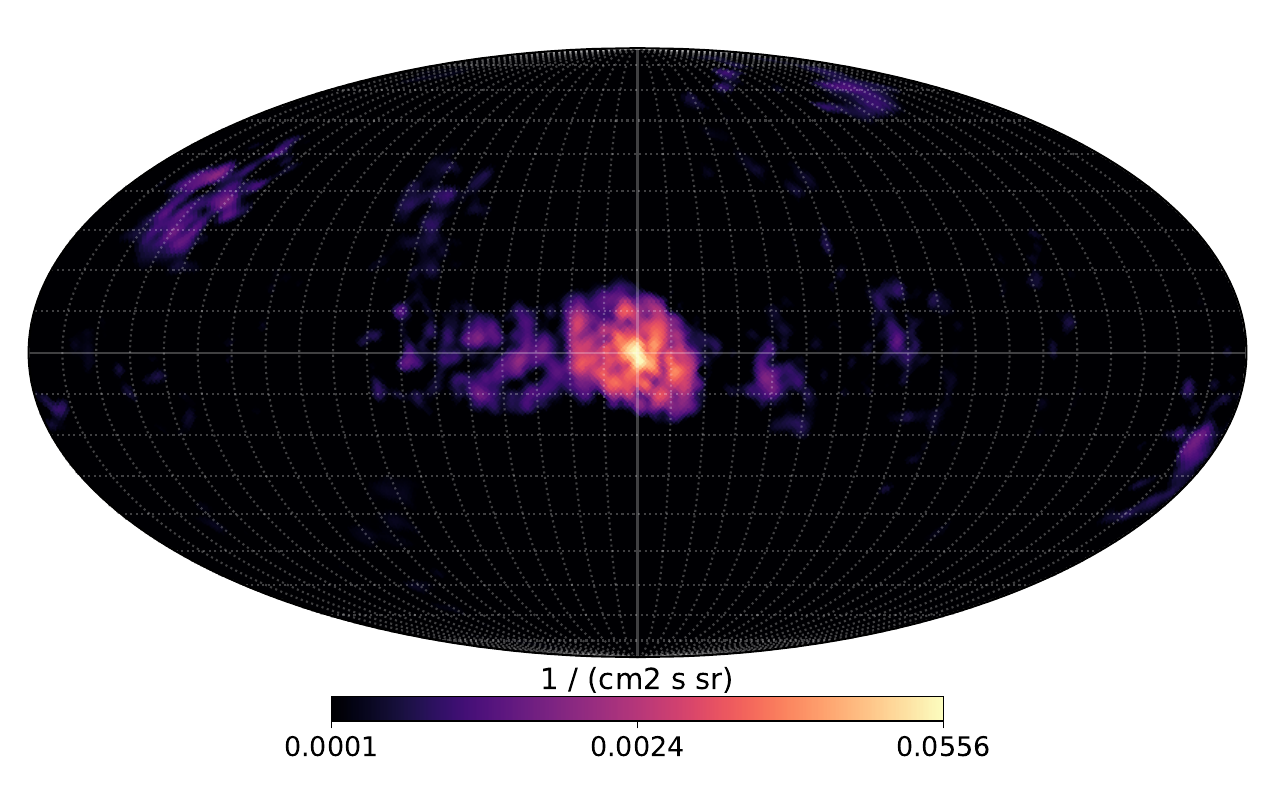}
    \includegraphics[width=0.32\linewidth]{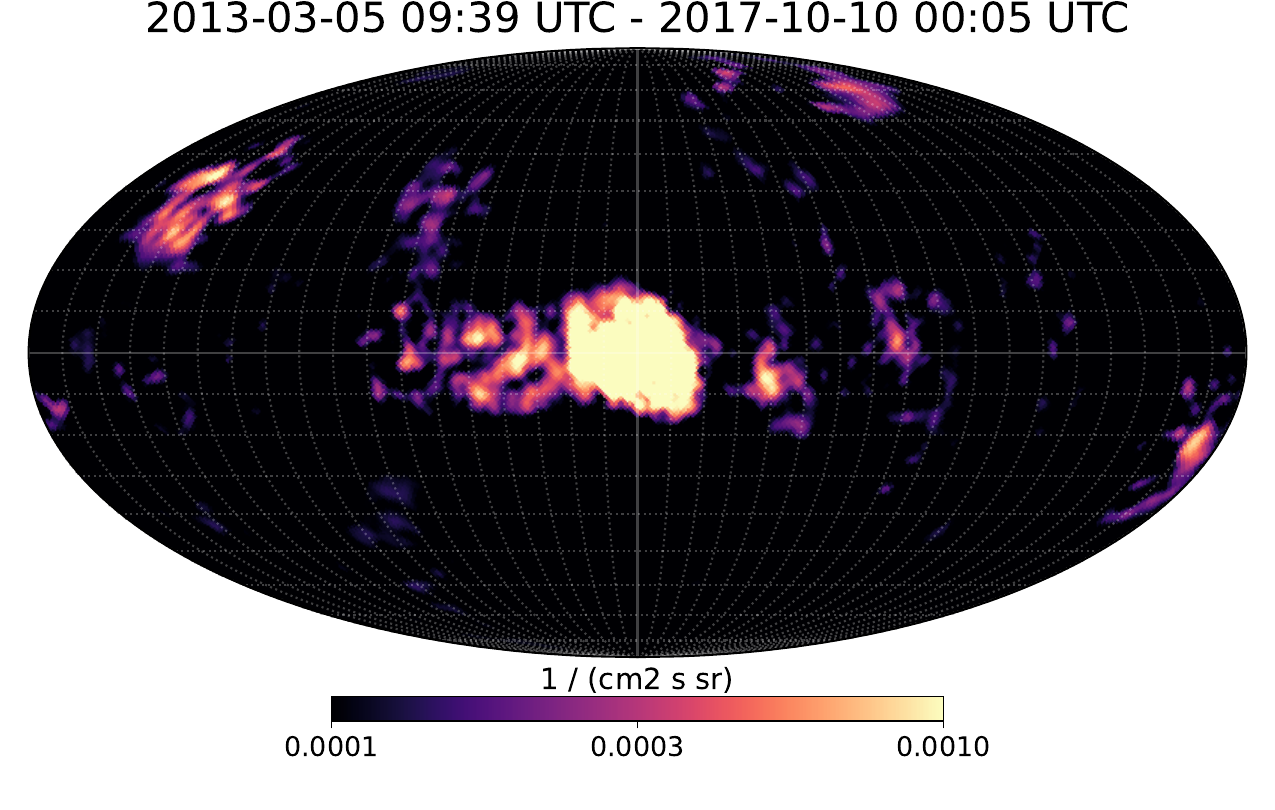}
    \includegraphics[width=0.32\linewidth]{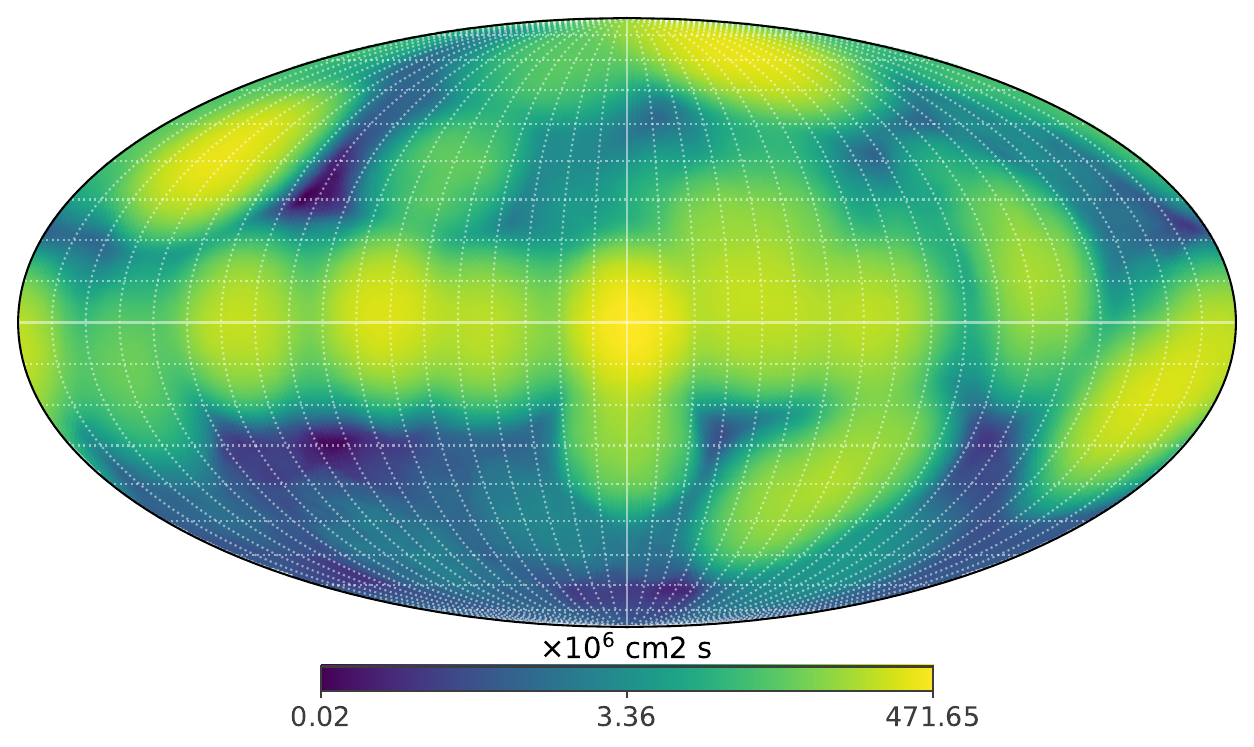}
    \includegraphics[width=0.32\linewidth]{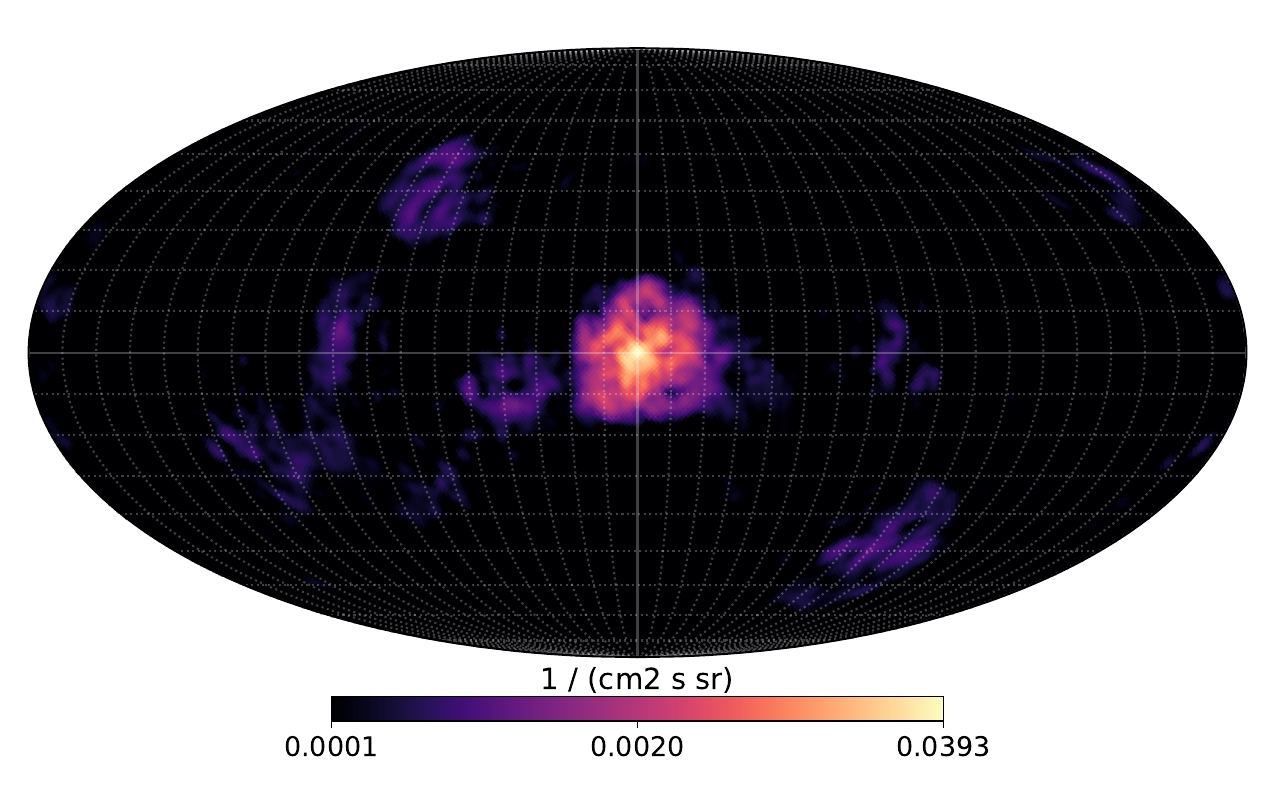}
    \includegraphics[width=0.32\linewidth]{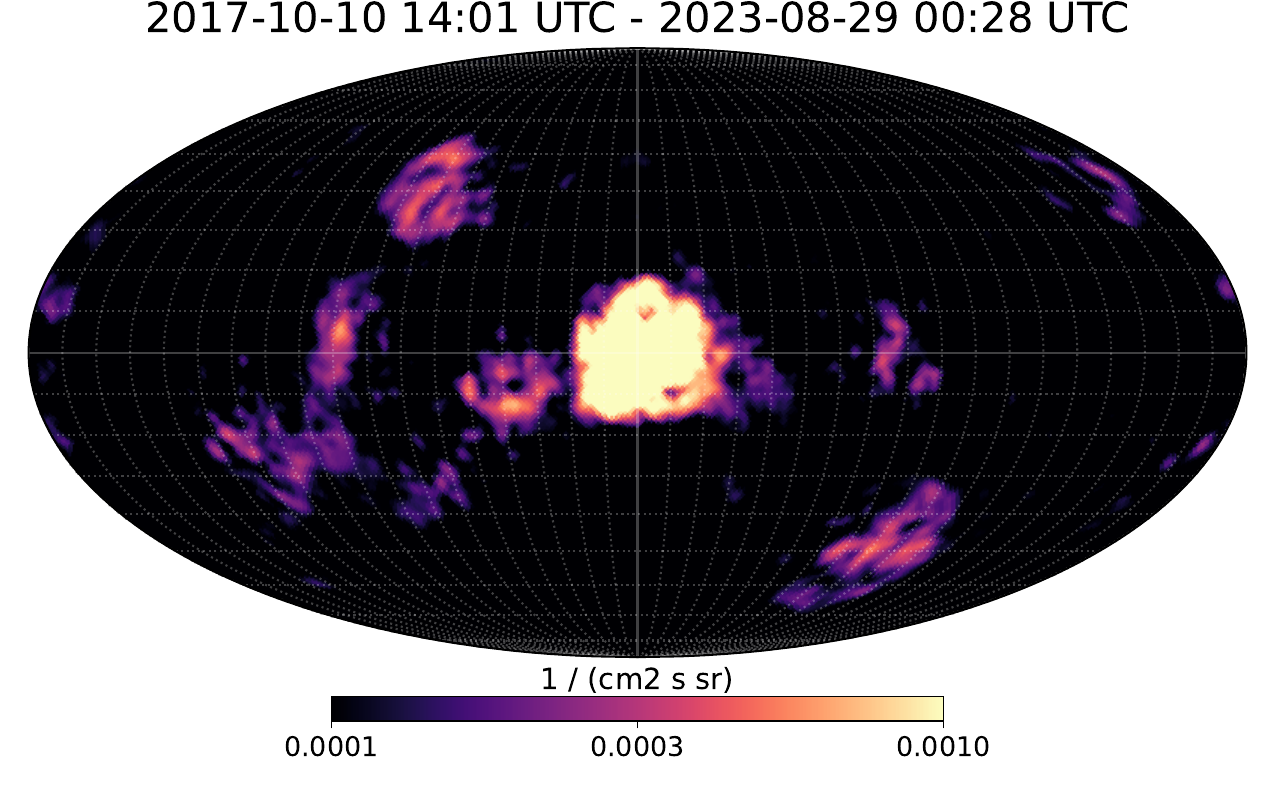}
    \includegraphics[width=0.32\linewidth]{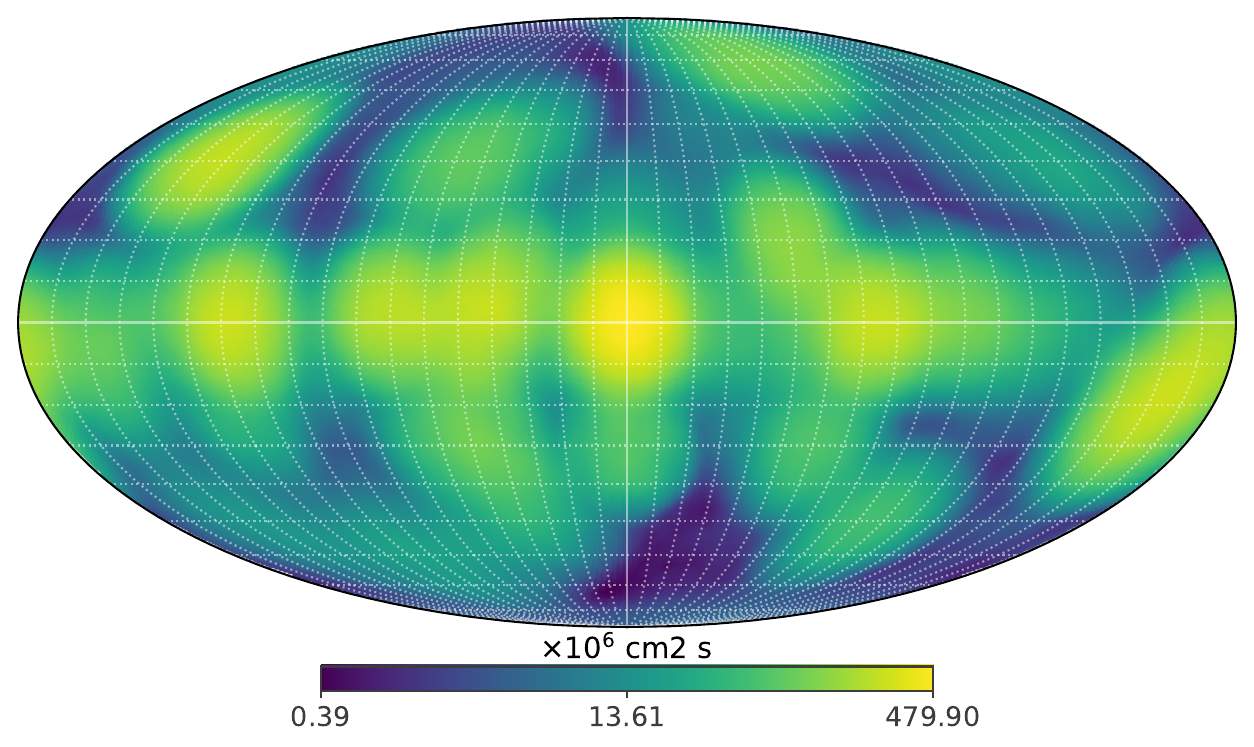}
    \caption{Reconstructed images using different independent periods of the data (from top to bottom: 2003-2008, 2008-2013, 2013-2017, and 2017-2023) with their corresponding exposure maps (right panels). In the middle panels, the maximum value is reduced to $1 \times 10^{-3}$ ph cm$^{-2}$ s$^{-1}$ sr$^{-1}$ to enhance the visibility of faint structures similar to Fig.~\ref{fig_bestimage_w_max1e-3}.}
    \label{fig_split_analysis}
\end{figure*}

\begin{figure*}
    \centering
    \includegraphics[width=0.32\linewidth]{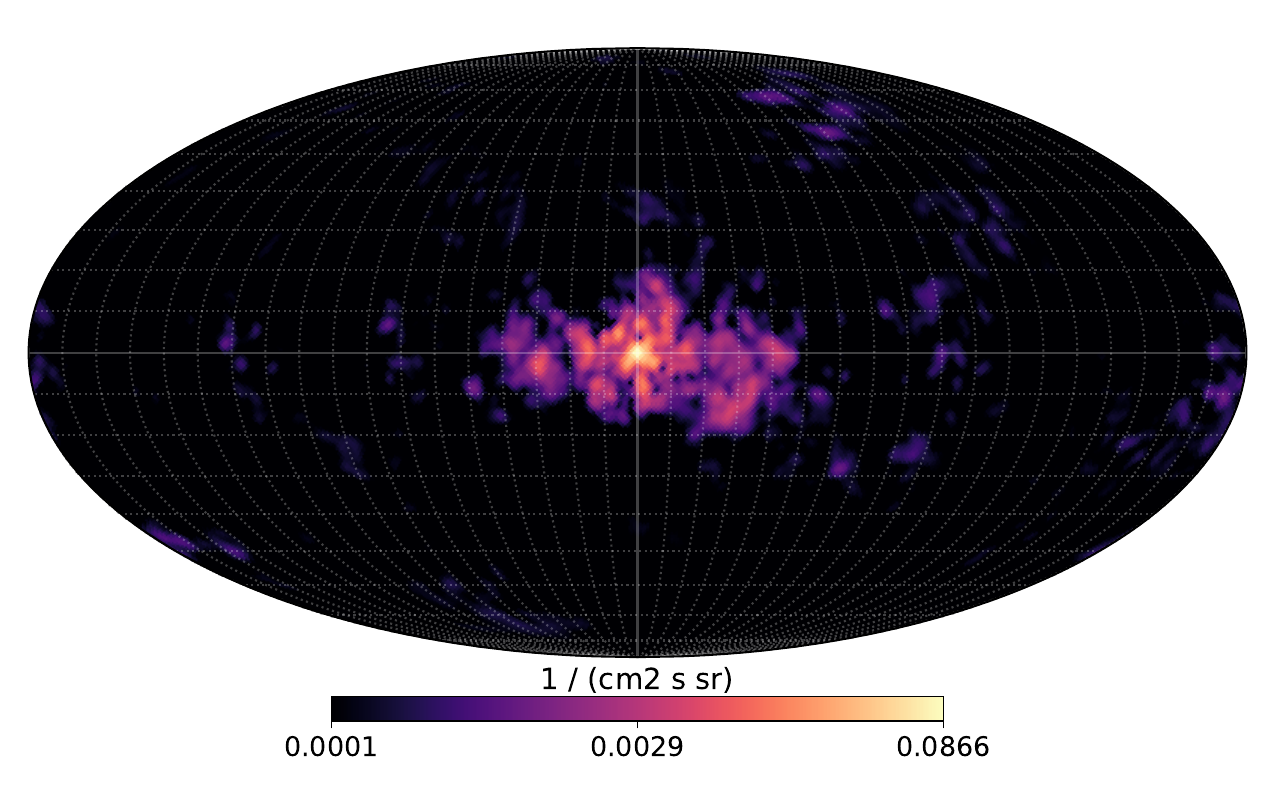}
    \includegraphics[width=0.32\linewidth]{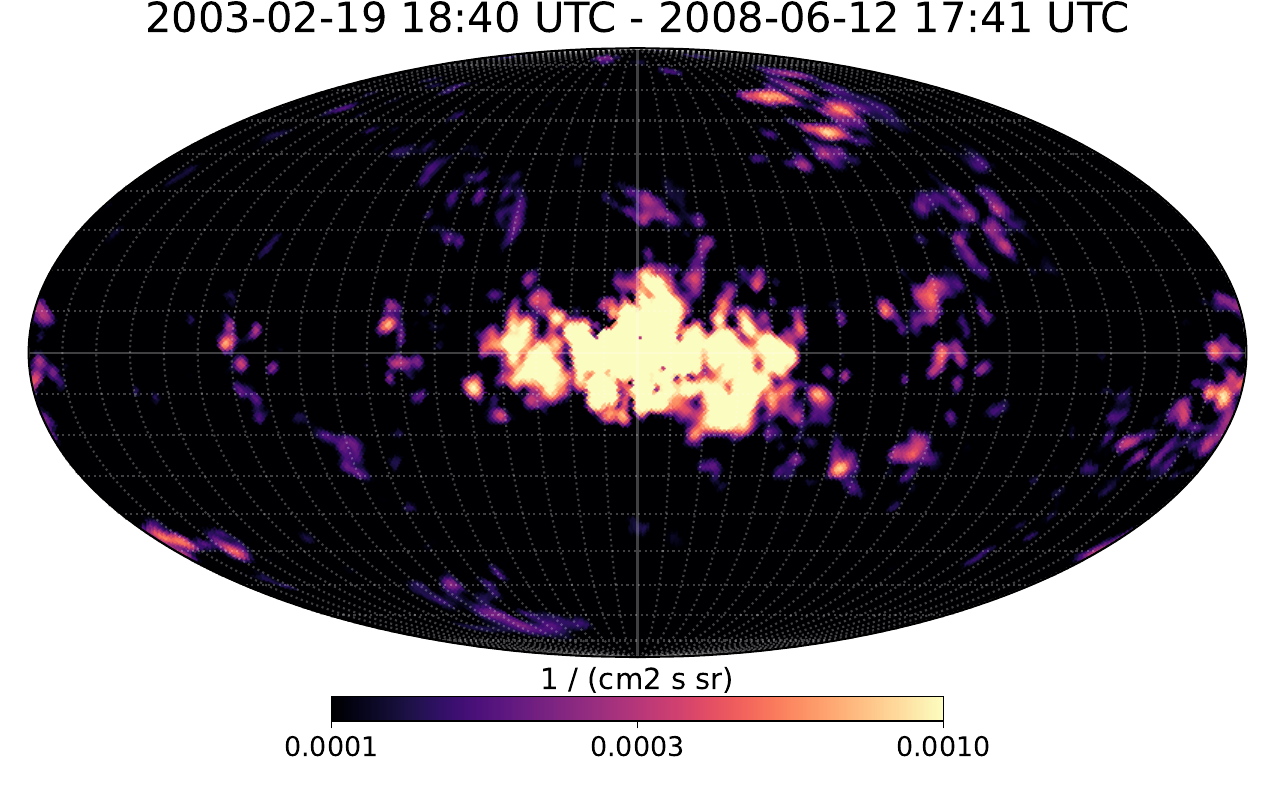}
    \includegraphics[width=0.32\linewidth]{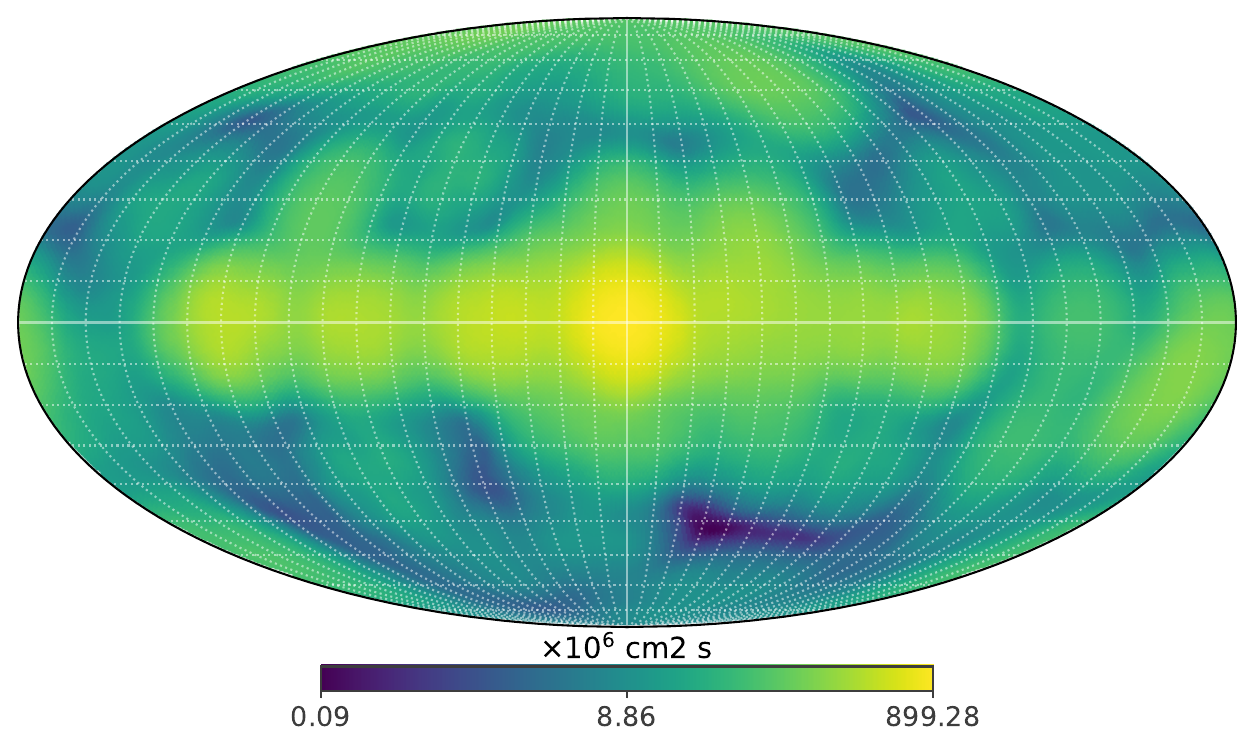}
    \includegraphics[width=0.32\linewidth]{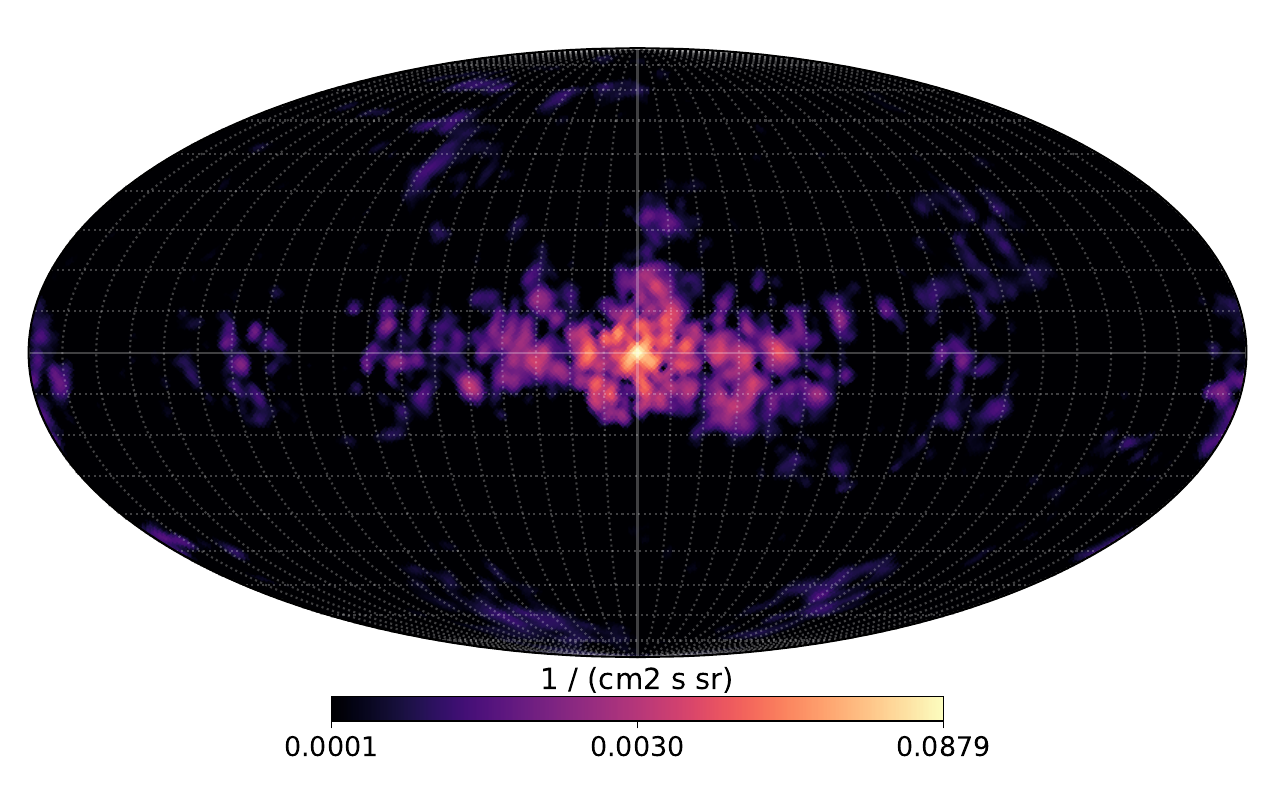}
    \includegraphics[width=0.32\linewidth]{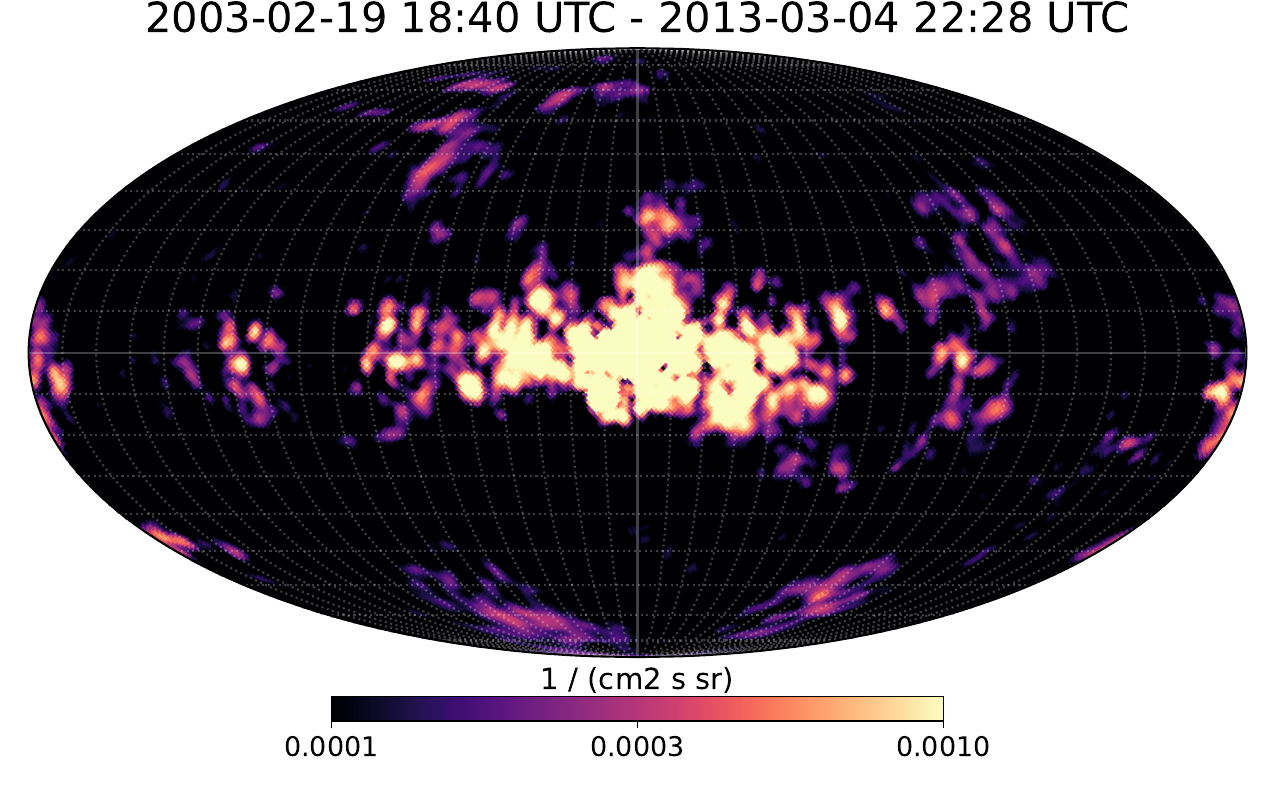}
    \includegraphics[width=0.32\linewidth]{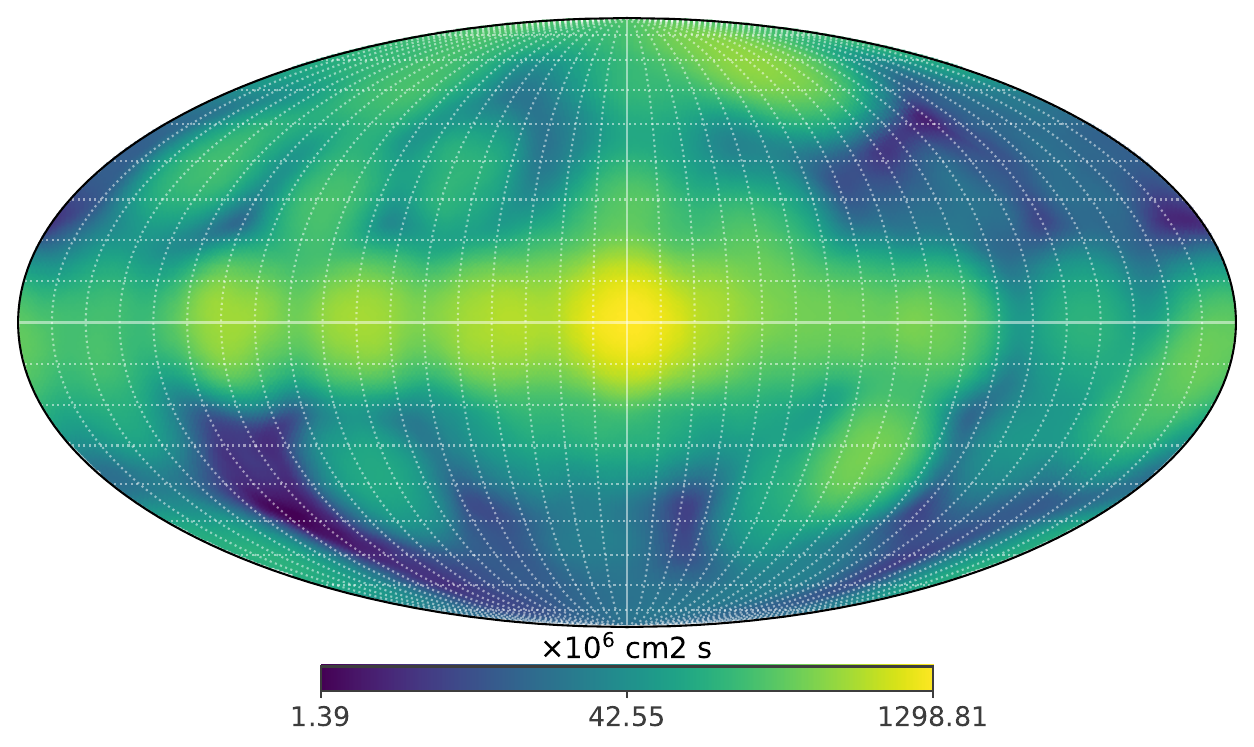}
    \includegraphics[width=0.32\linewidth]{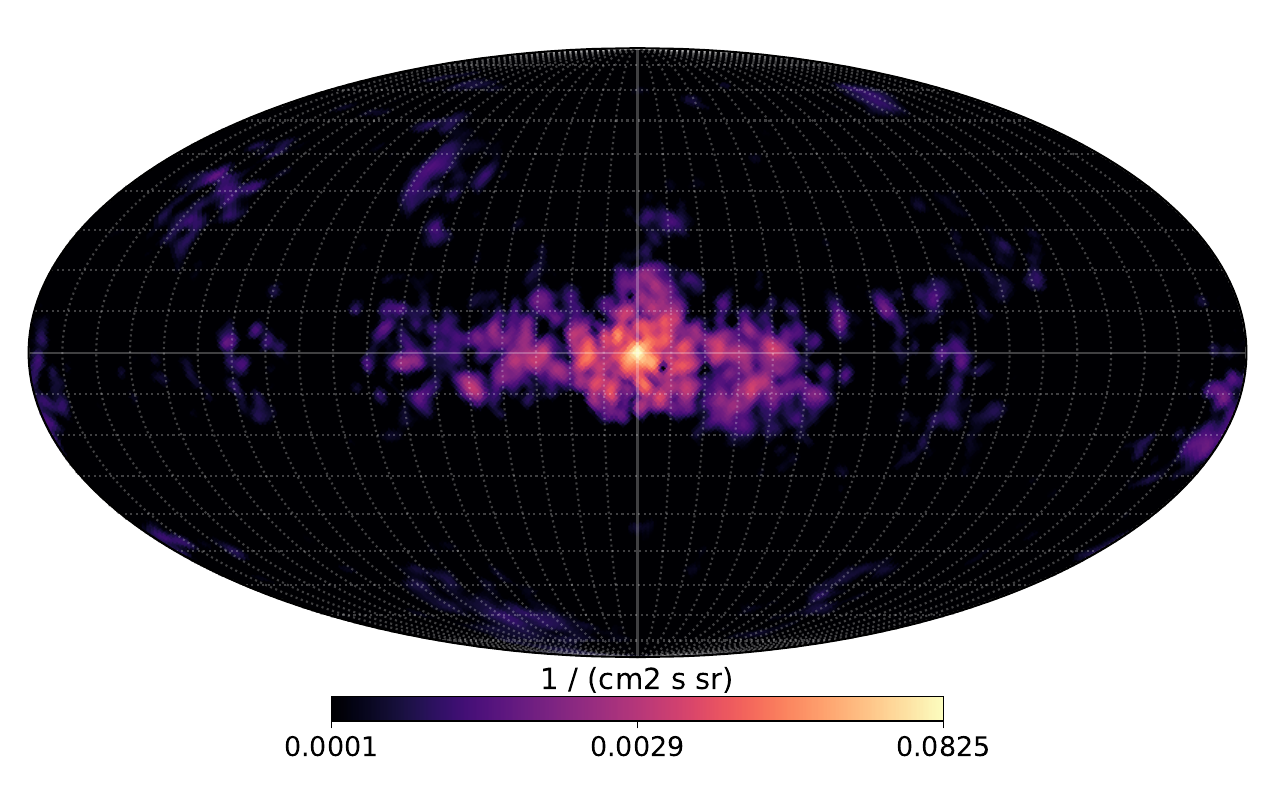}
    \includegraphics[width=0.32\linewidth]{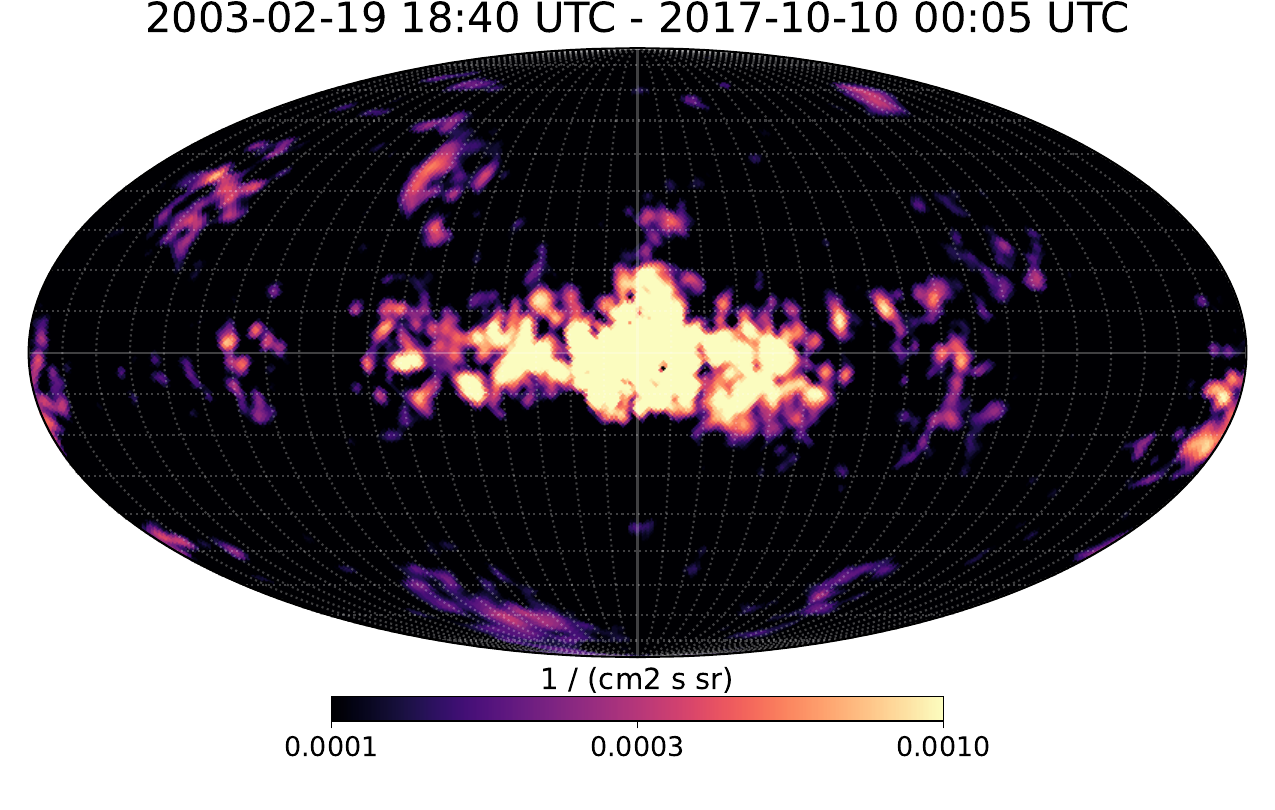}
    \includegraphics[width=0.32\linewidth]{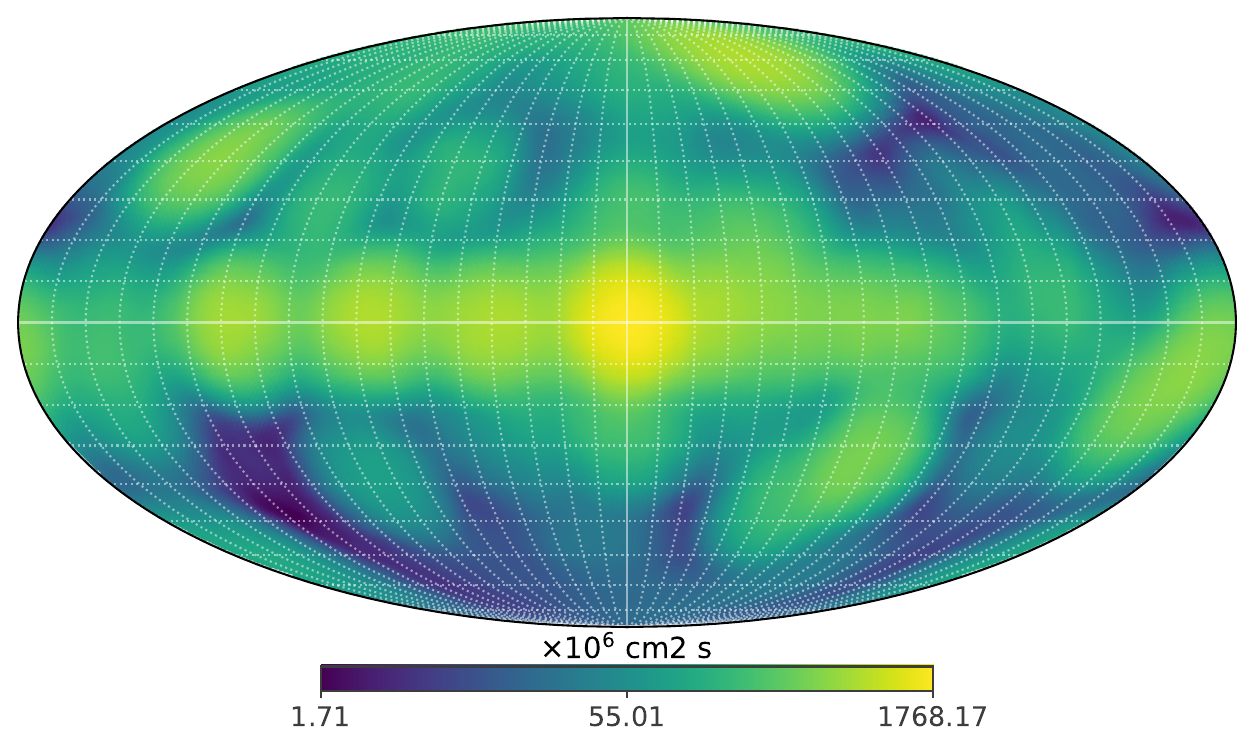}
    \includegraphics[width=0.32\linewidth]{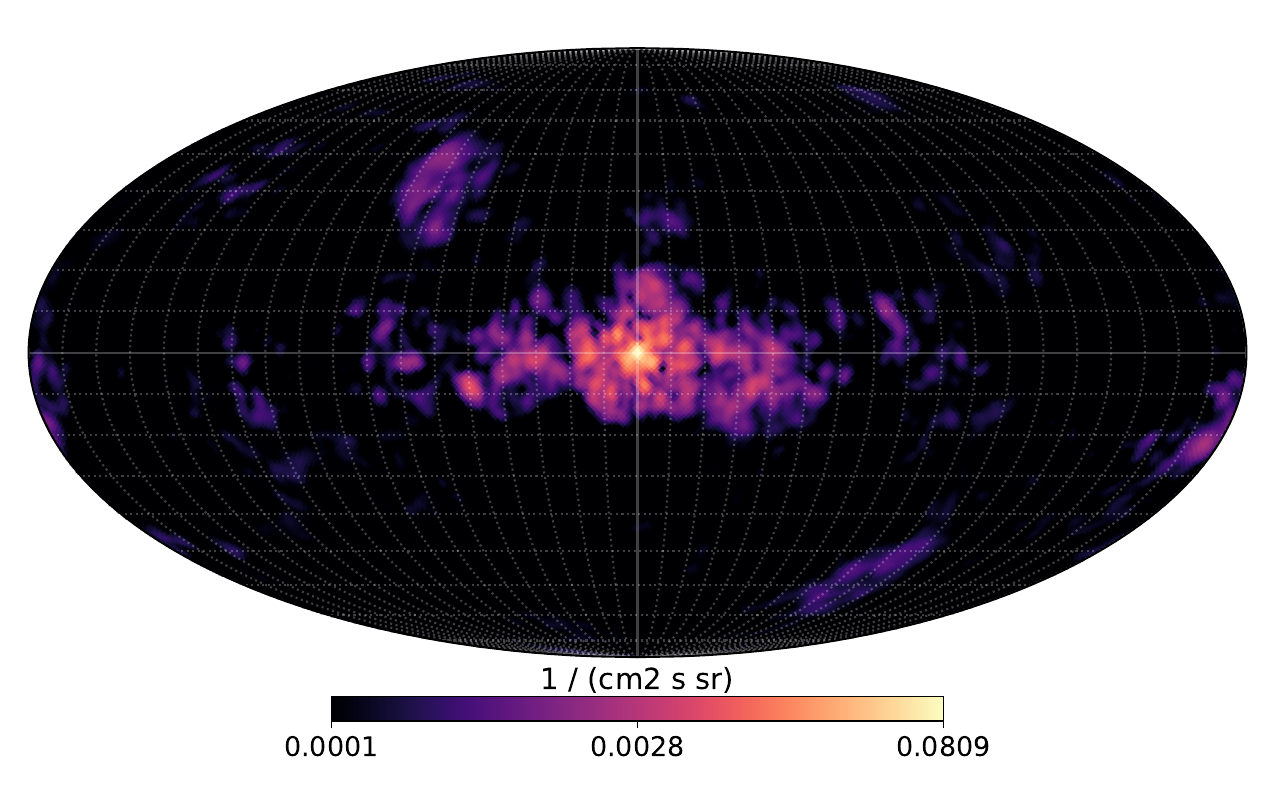}
    \includegraphics[width=0.32\linewidth]{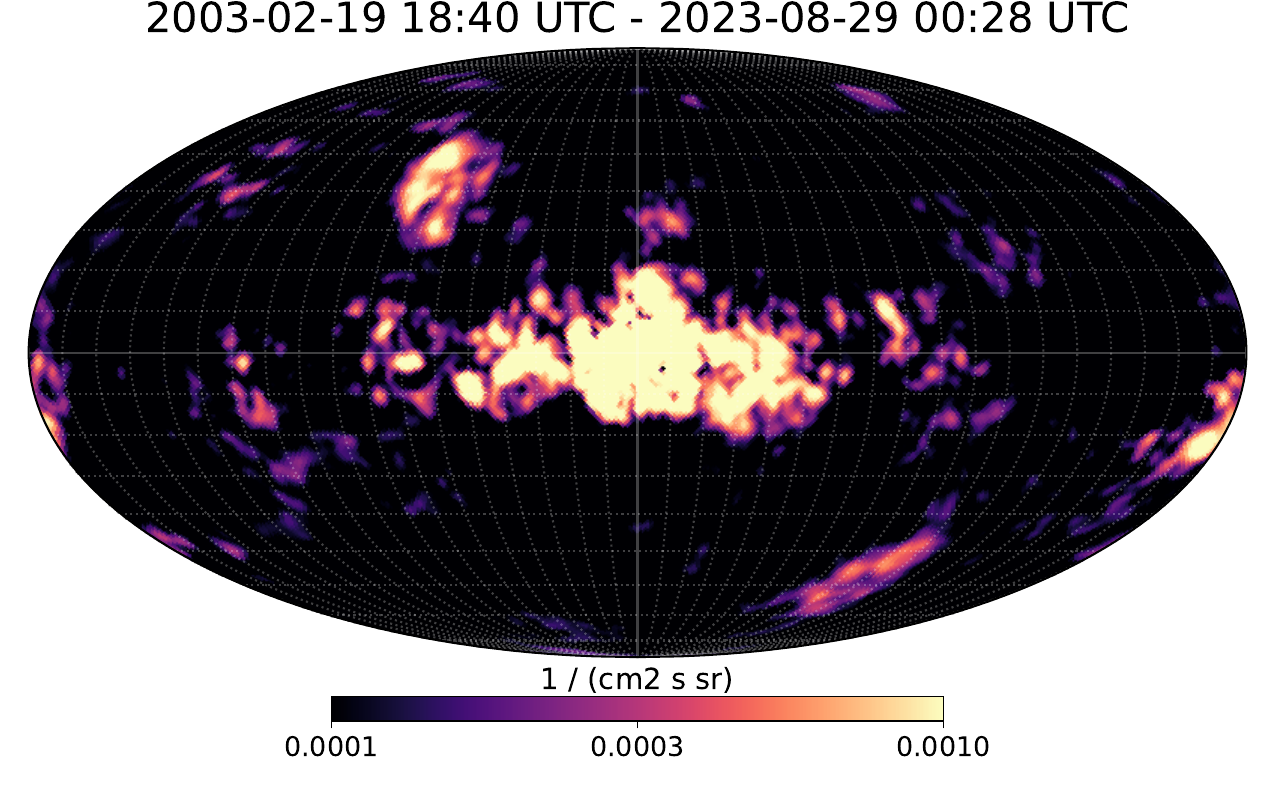}
    \includegraphics[width=0.32\linewidth]{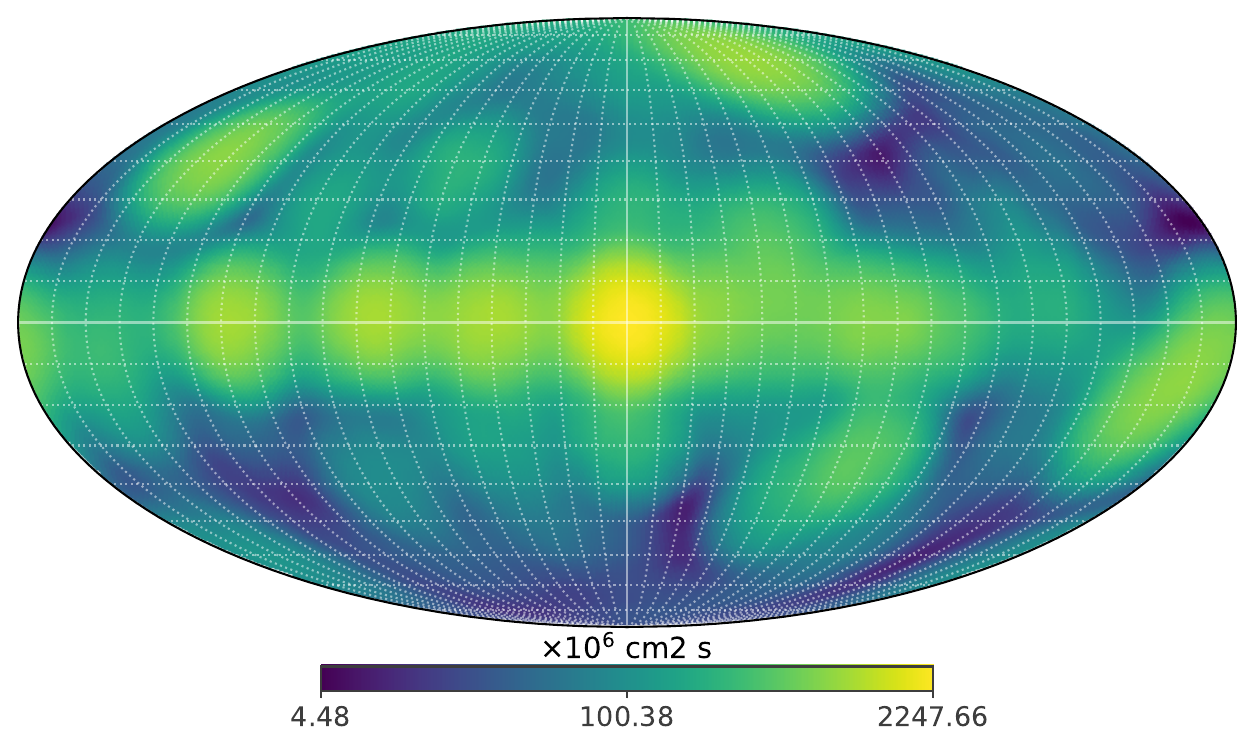}
    \caption{Same as Fig.~\ref{fig_split_analysis} but using different periods starting from 2003 with increasing durations (from top to bottom: 2003-2008, 2003-2013, 2003-2017, and the full dataset).}
    \label{fig_cumulative_analysis}
\end{figure*}

We performed image reconstruction using different periods to investigate the time variability/stability of the detected features.
Figure~\ref{fig_split_analysis} shows the reconstructed images using four independent periods (2003-2008, 2008-2013, 2013-2017, and 2017-2023) with their corresponding exposure maps.
While the bulge and disk features are consistently detected in these images, some features, such as the high-latitude blobs, appeared predominantly in specific periods when exposure close to these regions was long, but they are on the edge of the deep exposure regions.
It may suggest a potential systematic effect due to the combination of non-uniform exposure and the uncertainties of the off-axis response matrix.
As another example, the third image shows the bright region around $(l,b) \approx (140^{\circ}, 40^{\circ})$, where INTEGRAL performed deep observations of $\sim 4$ Ms for the type Ia supernova SN2014J \citep{Churazov2014,Diehl2015}.
We also show cumulative images starting from 2023 with increasing durations in Fig.~\ref{fig_cumulative_analysis}.

\section{Bootstrap samples for individual regions}
\label{sec_bootstrap_figures}

\begin{figure*}
    \centering
    \includegraphics[width=0.4\linewidth]{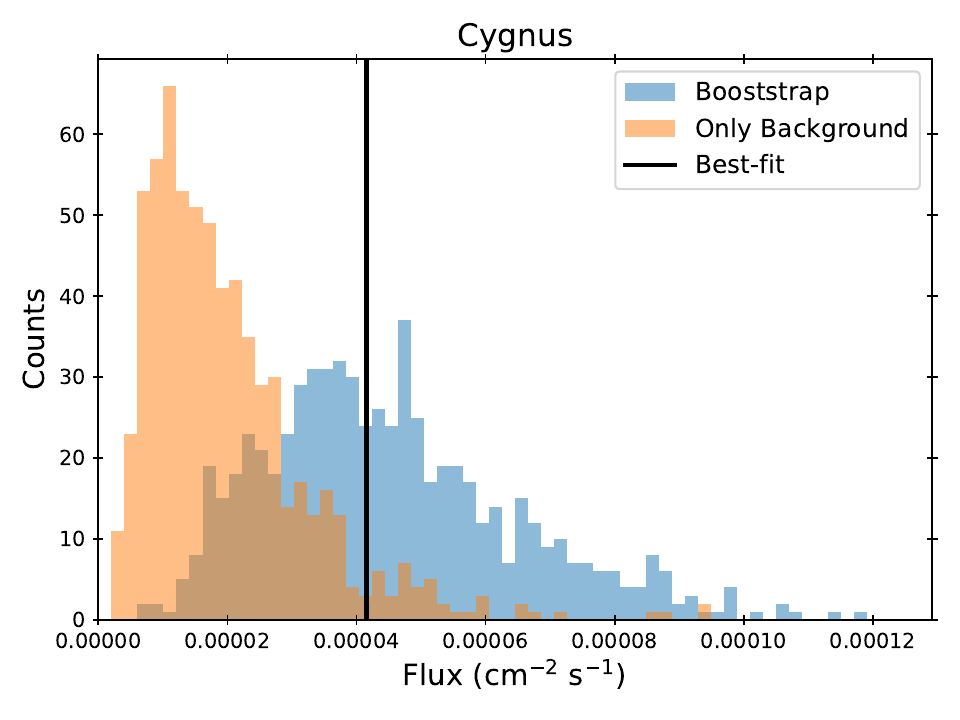}
    \includegraphics[width=0.4\linewidth]{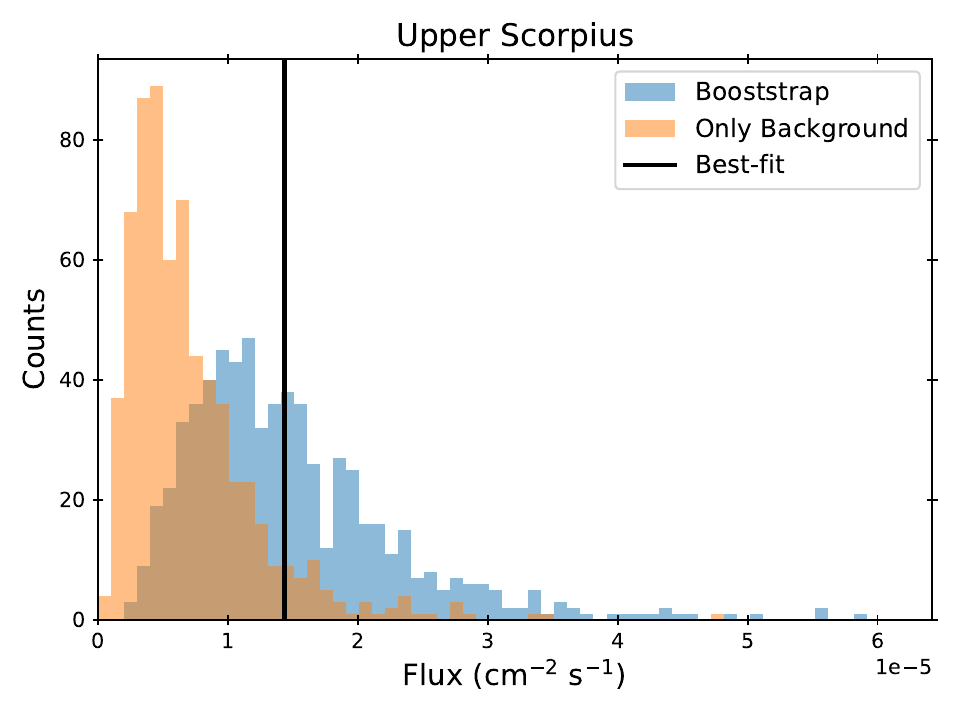}
    \includegraphics[width=0.4\linewidth]{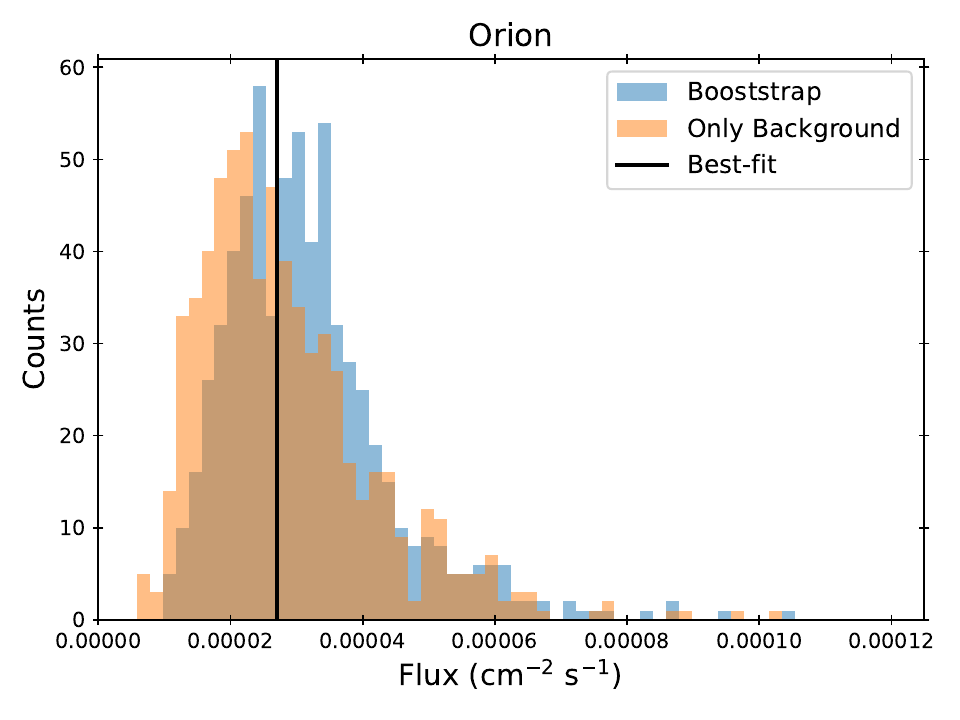}
    \includegraphics[width=0.4\linewidth]{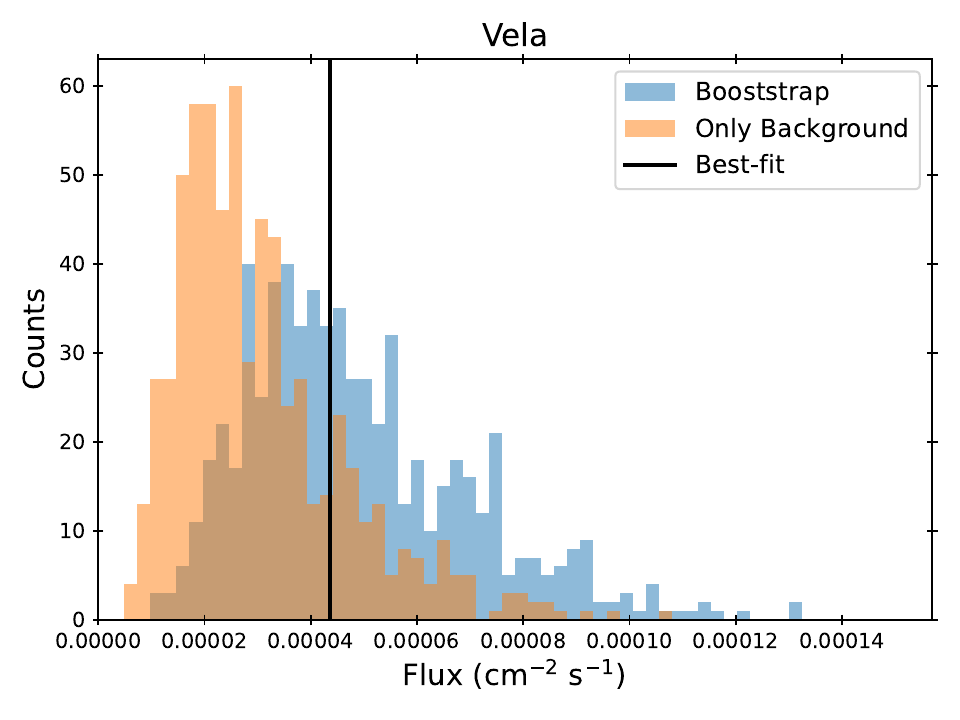}
    \caption{Flux distributions for individual regions derived from bootstrap samples. The Cygnus (top left), Upper Scorpius (top right), Orion (bottom left), and Vela (bottom right) regions are shown here.}
    \label{fig_flux_dist}
\end{figure*}

Figure~\ref{fig_flux_dist} shows the flux distributions for individual regions derived from bootstrap samples discussed in Sect.~\ref{sec_search_for_511keV_from_SFR}.
The blue histograms show the distributions of bootstrap samples, which were resampled from the data by adding Poisson noise. The orange histograms show those from the background-only samples. The vertical lines indicate the measured flux values from the reconstructed image shown in Fig.~\ref{fig_best_image}. 

\end{appendix}

\end{document}